\documentclass[journal]{IEEEtran}

\usepackage{textcomp}
\usepackage{cite}
\usepackage{amssymb,amsmath}
\usepackage{mathtools}
\usepackage{amsfonts, color}
\usepackage{multicol}
\usepackage{graphicx}
\usepackage{url}
\usepackage{modamsthm}
\usepackage{algorithm}
\usepackage{algorithmicx}
\usepackage{algpseudocode}
\usepackage{graphics,graphicx}
\usepackage{epstopdf , epsfig}
\usepackage{psfrag}
\usepackage{xcolor}
\usepackage{tikz}
\usepackage{pgfplots}\pgfplotsset{compat=newest}
\usepgfplotslibrary{fillbetween}
\usepgfplotslibrary{groupplots,units}
\usetikzlibrary{patterns}
\usetikzlibrary{shapes,arrows,shadows,positioning,decorations.pathreplacing,decorations.text,trees,calc}
\usepackage{pgfplots}
\usepackage[caption=false,font=footnotesize]{subfig}
\DeclareGraphicsExtensions{.eps}

\usepackage{latexsym}
\usepackage{slashed}
\usepackage{centernot}

\def\endthebibliography{%
  \def\@noitemerr{\@latex@warning{Empty `thebibliography' environment}}%
  \endlist
}

\begin{document}

\title{Interference Mitigation for Ultrareliable Low-Latency Wireless Communication}



\author{S. Arvin~Ayoughi, Wei~Yu,~\IEEEmembership{Fellow,~IEEE} Saeed R. Khosravirad, and Harish Viswanathan,~\IEEEmembership{Fellow,~IEEE}

\thanks{This work was supported in part by the Nokia--Bell Labs and in part by the Natural Sciences and Engineering Research Council (NSERC) of Canada.

S. A. Ayoughi and W. Yu are with The Edward S. Rogers Sr. Department of
Electrical and Computer Engineering, University of Toronto, 10 King's
College Road, Toronto, Ontario M5S 3G4, Canada (e-mails: sa.ayoughi@mail.utoronto.ca; weiyu@comm.utoronto.ca)

S. R. Khosravirad and H. Viswanathan are with Nokia--Bell Labs, Murry Hill, NJ 07974-0636 usa (e-mails: saeed.khosravirad@nokia-bell-labs.com; harish.viswanathan@nokia-bell-labs.com)}}

\maketitle

\begin{abstract} %

This paper proposes interference mitigation techniques for provisioning ultrareliable low-latency wireless communication in an industrial automation setting, where multiple transmissions from controllers to actuators interfere with each other. Channel fading and interference are key impairments in wireless communication. This paper leverages the recently proposed ``Occupy CoW'' protocol that efficiently exploits the broadcast opportunity and spatial diversity through a two-hop cooperative communication strategy among distributed receivers to combat deep fading, but points out that because this protocol avoids interference by frequency division orthogonal transmission, it is not scalable in terms of bandwidth required for achieving ultrareliability, when multiple controllers simultaneously communicate with multiple actuators (akin to the downlink of a multicell network). The main observation of this paper is that full frequency reuse in the first phase, together with successive decoding and cancellation of interference, can improve the performance of this strategy notably. We propose two protocols depending on whether interference cancellation or avoidance is implemented in the second phase, and show that both outperform Occupy CoW in terms of the required bandwidth and power for achieving ultrareliability at practical values of the transmit power.



















\end{abstract}


\maketitle

\section{Introduction}
\label{sec_intro}

\subsection{Background}
\label{sec_background}


\IEEEPARstart{W}{ireless} communication technologies enable a virtually ubiquitous access to information in a communication network. This facilitates mobility of transmitter and receiver entities. In an automated industrial factory, where sensors, controllers, and actuators continuously exchange control information, it is desirable to reduce cable installations and facilitate mobility of the devices. Wireless connectivity of machines enables a more flexible production of goods in factories. The intent of communicating control information among devices is to stabilize remote processes by feedback, which requires a near-real-time and highly reliable continuous streaming of information. Traditionally, the stringent requirements of latency and reliability are realized over wired
networks, see, e.g., \cite{fieldbus, ethernet, survey}. This paper studies the feasibility of providing ultrareliable connectivity through \emph{wireless} communications.



Using wireless networks in industrial communication for control applications introduces new challenges for wireless physical layer design. The current wireless cellular networks are designed for providing a high long-term average user rate for supporting a quality of service that is suitable for human visual and auditory perception. For ultrareliable low-latency machine-type communication, however, new protocols are needed to combat the impairments of the wireless channel, i.e., fading and interference, in order to meet the stringent requirements in the worst case of one-shot channel uses; see, \cite{TRS, requsandsols}. In this regard, diversity techniques \cite{sdiv, oc15, oc17} have been proposed to combat channel fading. This paper focuses on the equally important issue of interference mitigation. 

\subsection{Occupy CoW Protocol}

To combat fading, it is crucial to exploit diversity. In a rich-scattered environment, the distributed antennas of multiple devices in the network experience independent small-scale
fading, hence are an available and reliable source of spatial
diversity, which can be exploited by using a suitable cooperative
communication scheme \cite{sdiv, oc15, oc17}. The approach proposed in
\cite{oc17}, called ``Occupy CoW'', exploits this type of diversity by concatenating the messages for multiple devices together, then broadcasting the concatenated message, and further allowing cooperative communication to take advantage of diversity to combat fading. In the broadcast phase, the controller concatenates all the control messages of all actuators into one message, and exploits the multiuser diversity in broadcasting the concatenated message to all of the actuators. An actuator decodes this message if its wireless channel is not in a deep fade and is strong enough to support the rate of the concatenated message. In the cooperation phase, the cooperative diversity among
nodes is exploited and the message is relayed by all nodes that have successfully decoded in the first phase for the rest of the actuators. 
The key technique in this cooperation phase is the distributed space-time coding \cite{dstc}. In this phase, relays encode the concatenated message into a distributed linear dispersion code in the frequency domain and simultaneously transmit to obtain full diversity gain at the receivers. 

\subsection{Motivation and Main Contributions}
\label{sec_motivation}

This paper focuses on the mitigation of mutual interference with
multiple diversity transmissions in a wireless network. We ask the
following question: When multiple interfering controllers broadcast messages to their associated actuators, how should we combat fading and mitigate interference at the same time in order to enable ultrareliable low-latency wireless communication? 


The Occupy CoW protocol \cite{oc17} avoids interference by orthogonalization in the frequency domain. 
When multiple controllers are present in the network, each is assigned
a different frequency band. Hence, the interference among the
distributed transmitters is avoided; this orthogonalization approach
further enables cooperation among the half-duplex relaying nodes in the second
phase for exploiting spatial diversity. But, the main drawback is
that the required bandwidth needs to scale linearly with the network size.

A candidate scheme for improving the scalability of the protocol with
number of transmitters is full frequency reuse---instead of orthogonal
transmission by frequency division. However, we observe through
simulations that if interference is treated as noise, it would have a severe adverse effect on the failure probability. The impact of treating interference as noise is most notable in the cooperation phase of the protocol, where all relays transmit. Therefore, the
effect of interference challenges the practicality of the Occupy CoW
protocol in large networks at reasonable amounts of bandwidth and
transmit power.

In this paper, we propose two protocols for improving the scalability
of Occupy CoW in large networks. The first protocol consists of full
bandwidth reuse by both the controllers and actuators, while allowing
successive decoding and cancellation of interference at actuators in
both broadcast and cooperation phases. The weak interference signals
that remain undecoded are treated as noise. We observe that this
scheme successfully improves the scalability and outperforms other
protocols in the low-power regime. But, in the high-power regime, it
suffers from a failure probability floor and a low diversity order. The second protocol removes these shortcomings, while still preserving
the improvement in scalability to a large extent. In the broadcast
phase the second protocol still uses the successive interference
decoding and cancellation scheme for improving scalability. In the
cooperation phase, it uses frequency orthogonalization to avoid
interference and to exploit spatial diversity.

We identify regions in the failure probability versus transmit power
plane in which the two proposed interference mitigation protocols and
the original Occupy CoW protocol may outperform each other. In terms
of bandwidth consumption, our first protocol is preferable in the
low-power regime; the Occupy CoW protocol is preferable in the
high-power and low-bandwidth regime; our second protocol is the
preferred scheme in the medium-power regime.

In \cite{oc17}, the Occupy CoW protocol is analyzed in a single wireless local
domain where all nodes are in range of each other and wireless channel
gains are i.i.d.\ across the network. The distance-dependent path loss of the wireless channel is not considered. 
In this paper, we consider more realistic distance-dependent wireless 
channel models. Our conclusions are obtained using Monte Carlo simulations 
of failure probabilities at varying network parameters.

\subsection{Other Related Works}

To reduce the required transmit power for achieving ultrareliability
by Occupy CoW protocol, \cite{ncoding} proposes the XOR-CoW protocol,
which uses a network coding approach for combining uplink and
downlink transmissions to improve the use of bandwidth. 

Adaptive relay selection is proposed in \cite{relsel} to exploit cooperative diversity only from those relays that have a strong source-relay and relay-destination channel condition. A well-designed relay selection scheme reduces the total power consumption in the network for achieving ultrareliability. It simplifies practical implementation of simultaneous relaying for exploiting cooperative diversity. Moreover, reducing the number of active relays can improve reliability by reducing the interference. However, any relay selection strategy requires a certain level of channel state information at the transmitter. Reliable relay selection in large networks with limited channel state information at the transmitter side is challenging. 

For mitigating interference and improving diversity order,
\cite{deploymenturll} suggests increasing the number of antennas at nodes.
Further, it proposes having partial frequency reuse in the network, i.e.,
having reuse factor of 1 for cell-center receivers and less than 1 for
cell-edge receivers, to improve spectral efficiency. Also, \cite{relayselection} proposes installing extra multi-antenna stationary relay nodes in the network and uses a relay selection scheme to combat fading. It is assumed that transmitters
have full knowledge of channels and use it for selecting the best relay.
In our protocols, the relaying scheme is distributed and channel state
information is not needed at transmitters.

The idea of successive intra-cell interference cancellation is proposed and studied in various scenarios of wireless cellular networks. In \cite{nomaforfairness}, it is shown that non-orthogonal access with successive intra-cell interference cancellation improves the cell-edge user rates compared to orthogonal access. Also, in \cite{noma13} it is observed that non-orthogonal multiple access with successive interference cancellation improves both the capacity and the cell-edge rates. In this work, however, we deal with inter-controller interference mitigation for ultrareliable low-latency wireless communication. In this case, especially for a cell-edge actuator, when the channel condition from interfering controllers is better than the channel condition from the intended controller, successive cancellation of inter-cell interference is possible. 






\subsection{Organization of the Paper}

Section \ref{sec_sys} presents the system model. We  introduce three benchmark interference avoidance Occupy CoW-based protocols in Section \ref{sec_avoidance}. We analyze these protocols in Section \ref{sec_analysis}. The proposed protocol with interference cancellation in both phases is presented in Section \ref{sec_sic}. The proposed protocol with interference cancellation in the first phase and interference avoidance in the second phase is presented in Section \ref{sec_icia}. Section \ref{sec_sim} presents results of Monte Carlo simulations. Finally, Section \ref{sec_con} concludes the paper.

\section{System Model}
\label{sec_sys}

We consider an industrial factory hall in which a number of automated production lines are deployed close to each other. In each production line, a closed-loop controller stabilizes a number of remote processes over wireless channels. This requires continuous streaming of information from sensors to the controller and from the controller to actuators. The focus of this work is on the controller-to-actuator communication. Using the terminology of the wireless cellular networks literature, we refer to the area occupied by a production line as a cell and to transmission of the control messages from controller to actuators as the downlink transmission.

We denote the total number of interfering cells (or controllers) in the factory hall by $C$, and the number of actuators in each cell by $K$. The set of all controllers is $\mathcal{C} = \{1, \dots, C\}$. The $k$th actuator in cell $c \in \mathcal{C}$ is referred to as actuator $(c, k)$, and $\mathcal{K} = \mathcal{C} \times \{1, \dots, K \}$ is the set of all actuators.

In each cell, the controller has $K$ independent messages of $b$ bits, one for each actuator. All messages are required to be communicated within $T$ seconds, over the total available bandwidth of $W$ Hz.

We allow two-hop cooperative communications for exploiting spatial diversity of distributed antennas of actuators. The two hops are orthogonal in time, i.e., they take place consecutively. The first hop is the broadcast hop, which is the transmission from the controller to the actuators, and the second one is the cooperation hop, which is the cooperative communication among the half-duplex actuators. 

In the broadcast phase, actuator $(c, k)$ receives
\begin{align}
Y_{c, k}^{\mathrm{(I)}}(t) = \sum_{i = 1}^C g_{c, k, i}  X_i^{\mathrm{(I)}}(t) + N_{c, k}^{\mathrm{(I)}}(t),
\label{chn_1}
\end{align}
for $t \in (0, T/2)$, where ${g}_{c, k, i} \in \mathbb{C}$ is the realization of channel gain ${G}_{c, k, i}$ from controller $i$ to actuator $(c, k)$ for the transmission period $t \in (0, T)$, $ X_i^{\mathrm{(I)}}(t)$ is the zero-mean Gaussian transmit signal from the $i$th controller over $t \in (0, T/2)$ with power spectral density $p$ dBm/Hz, $N_{c, k}^{\mathrm{(I)}}(t)$ is the background additive white Gaussian noise (AWGN) process with power spectral density $\sigma^2$ dBm/Hz.

In the cooperation phase, let $\mathcal{S}$ denote the set of actuators that would participate in cooperation in the second phase, and let $X_{c', k'}^{\mathrm{(II)}}(t)$ be the zero-mean Gaussian transmit signal from actuator $(c', k') \in \mathcal{S}$ ($X_{c', k'}^{\mathrm{(II)}}(t) = 0$ for $(c', k') \notin \mathcal{S}$), and $X_i^{\mathrm{(II)}}(t)$ be the zero-mean Gaussian transmit signal from the $i$th controller, all over $t \in (T/2, T)$ with power spectral density $p$ dBm/Hz. In this second phase, each actuator $(c, k)$ receives
\begin{multline}
Y_{c, k}^{\mathrm{(II)}}(t) = \sum_{i = 1}^C  g_{c, k, i} \bar{X}_i^{\mathrm{(II)}}(t) \\ + \sum_{(c', k') \in \mathcal{S}} {h}_{c, k, c', k'} \bar{X}_{c', k'}^{\mathrm{(II)}}(t) + N_{c, k}^{\mathrm{(II)}}(t).
\label{chn_2}
\end{multline}
Note that the actuators act as half-duplex distributed relays, i.e., they can simultaneously transmit and receive only if their transmission and reception are orthogonal in the frequency domain, so that they can only receive the components of $X_i^{\mathrm{(II)}}(t)$ and $X_{c', k'}^{\mathrm{(II)}}(t)$ that are orthogonal to $X_{c, k}^{\mathrm{(II)}}(t)$, denoted in (\ref{chn_2}) as $\bar{X}_i^{\mathrm{(II)}}(t)$ and $\bar{X}_{c', k'}^{\mathrm{(II)}}(t)$, respectively. Note that if $(c, k) \notin \mathcal{S}$, then $X_{c, k}^{\mathrm{(II)}}(t) = 0$ and we have $\bar{X}_i^{\mathrm{(II)}}(t) = X_i^{\mathrm{(II)}}(t)$ and $\bar{X}_{c', k'}^{\mathrm{(II)}}(t) = X_{c', k'}^{\mathrm{(II)}}(t)$. Here, ${h}_{c, k, c', k'} \in \mathbb{C}$ is the realization of channel gain ${H}_{c, k, c', k'}$ from actuator $(c', k')$ to actuator $(c, k)$ for the transmission period $t \in (T/2, T)$, $N_{c, k}^{\mathrm{(II)}}(t)$ is the background AWGN process with power spectral density $\sigma^2$ dBm/Hz. We define the signal-to-noise ratio (SNR) as
\begin{align}
\rho \triangleq \frac{p}{\sigma^2}.
\label{ratio}
\end{align}

Variations of channel gains in time or frequency can provide further diversity. To isolate the spatial diversity gain obtained by cooperation, channel gains are assumed to be constant over frequency and time for each transmission period, while varying from one transmission period to another, according to their fading probability distribution.

We assume availability of channel state information at receivers (CSIR) everywhere in the network. However, channel state information at transmitters is not needed. Moreover, we assume signals from multiple distributed transmitters can be synchronously combined using space-time coding. We note here that a certain level of asynchronism can provide delay diversity, as in the techniques of \cite{armin_cdd, armin_thesis}. For relevant synchronization techniques see, e.g., \cite{synchCoop, synchURLLC}.   

\section{Benchmark Protocols}
\label{sec_avoidance}

In this section, we introduce three variations of the previously-proposed Occupy CoW scheme for ultrareliable communication in a multicell scenario. We use these schemes as benchmarks in the rest of the paper. 

The Occupy CoW protocol avoids intra-cell interference in a cell by concatenating all independent messages into a single message to be broadcast to all receivers of that cell. In a multicell network, it avoids the inter-cell interference by orthogonalization. It is shown in the next section that as the SNR goes lower this orthogonalization approach requires an impractically large bandwidth for achieving ultrareliability in a multicell network.  

\subsection{Frequency Division Among Cells}
\label{sec_ooc}

First, let us consider the simple scheme of dividing the available bandwidth equally among $C$ cells and using the Occupy CoW protocol in each cell over the allocated bandwidth. We refer to this scheme as the Orthogonal Occupy CoWs (Orth-Occupy CoWs). We first review the protocol briefly.



In the $c$th cell, $c \in \mathcal{C}$, all $K$ independent messages are concatenated into one message of size $Kb$ bits at the controller, and all $K$ nodes in the cell are required to decode this concatenated message within two phases, each lasting $T/2$ seconds. Hence, the aggregate per cell rate in each phase amounts to
\begin{align}
R = \frac{K b}{0.5T} \  \ \ \mathrm{(b/s)}.
\label{rate}
\end{align}

In the first phase, the controller broadcasts the concatenated message to actuators. Only those actuators whose wireless channels are deeply faded and their link capacity is less than $R$ fail to decode the concatenated message. Therefore, in this phase, the set of successful actuators in cell $c$ that have decoded the concatenated message of controller $c$ is
\begin{align}
\mathcal{S}_c^{orth} = \left\{(c, k) \in \mathcal{K}: \frac{W}{C} \log \left(1 + \rho |g_{c, k, c}|^2\right) \geq R \right\}.
\label{first1c_o}
\end{align}

In the second phase, the controller retransmits, and all nodes in $\mathcal{S}_c^{orth}$ relay the message for the remaining $K - |\mathcal{S}_c^{orth}|$ receivers of the cell. To exploit the available spatial diversity, the message is transmitted using distributed space-time codes. 
The protocol fails if at least one of the actuators in the network fails to decode the message of its controller by the end of the second phase, i.e., if there is a $(c, k) \notin \mathcal{S}_c^{orth}$ for a $c \in \mathcal{C}$ for which
\begin{align}
\frac{W}{C} \log \left(1 + \rho |g_{c, k, c}|^2 + \sum_{(c', k') \in \mathcal{S}_c^{orth}} \rho |h_{c, k, c', k'}|^2\right) < R.
\label{second1c_o}
\end{align}
Here, obtaining spatial diversity by using distributed space-time coding is simply modeled by adding the received signal powers of all transmitters.

\subsection{Inter-Cell Cooperation Among Actuators}
\label{sec_oc}

In frequency division scheme of the last subsection, for each cell, the order of spatial diversity in the second phase is limited by the number of successful nodes of the first phase in that cell. This diversity order can be increased notably by allowing inter-cell cooperation among actuators of the network. This protocol is indeed the Occupy CoW scheme for generic information topology of \cite{oc17}. As we show in Section \ref{sec_analysis}, inter-cell cooperation among actuators can significantly improve the performance, indicating the importance of exploiting the available spatial diversity of actuators' antennas.


In this scheme, all actuators that have decoded the message of the $c$th controller in the first phase relay this message over its associated frequency band in the second phase. The set of these actuators is
\begin{align}
\mathcal{S}_c = \mathcal{S}_c^{orth} \cup \mathcal{S}_{c}^{coop},
\label{first1c_ooca}
\end{align}
where $\mathcal{S}_c^{orth}$ is the set of successful actuators in the $c$th cell that have decoded the message of the $c$th controller, as defined in (\ref{first1c_o}), and
\begin{multline}
\mathcal{S}_{c}^{coop} \\ = \left\{(c', k) \in \mathcal{K}: c' \in \mathcal{C}/ \{c\} , \frac{W}{C} \log \left(1 + \rho |g_{c', k, c}|^2\right) \geq R \right\}
\label{first1c_o_ooca}
\end{multline}
is the set of successful actuators outside of the $c$th cell that have decoded the $c$th controller's massage, and $R$ is defined in (\ref{rate}).

The protocol fails if at least one of the actuators in the network fails to decode the message by the end of the second phase, i.e., if there is a $(c, k) \notin \mathcal{S}^{orth}_c$ for a $c \in \mathcal{C}$ for which
\begin{align}
\frac{W}{C} \log \left(1 + \rho|g_{c, k, c}|^2 + \sum_{(c', k') \in \mathcal{S}_c} \rho |h_{c, k, c', k'}|^2\right) < R.
\label{second1c_ooca}
\end{align}

\subsection{Full Cooperation Among Cells}
\label{sec_foc}

In this section, we consider using the Occupy CoW protocol in the network with genie-aided message sharing among controllers. Message sharing among all $C$ controllers transforms the network into one large cell with $C$ distributed transmit antennas. This can be realized in practice, if controllers can be connected through high-speed backhaul links. Sharing messages among controllers improves the diversity order as compared to schemes of the last two subsections. We refer to this scheme as the Coordinated Multipoint Occupy CoW (CoMP-Occupy CoW).



In this protocol, all messages are concatenated into one message of size $CKb$ bits at the controllers, and all $CK$ actuators in the network are required to decode this concatenated message within two phases, each lasting $T/2$ seconds.

In the first phase, controllers broadcast the concatenated message to actuators, using distributed space-time codes for exploiting spatial diversity. Only those actuators whose wireless channels are in such a deep fade that their achievable rate is less than $CR$ fail to decode the concatenated message. Therefore, in this phase, the set of successful actuators is
\begin{align}
\mathcal{S} = \left\{(c,  k) \in \mathcal{K}: W \log\left(1+ \sum_{i \in \mathcal{C}} \rho | g_{c, k, i}|^2\right) \geq CR \right\}.
\label{first_fcoc}
\end{align}

In the second phase, the controllers retransmit, and all nodes in $\mathcal{S}$ relay the message for the rest of the receivers using distributed space-time coding. 
The protocol fails if at least one of the actuators fails to decode the message by the end of the second phase, i.e., if there is a $(c, k) \notin \mathcal{S}$ for which
\begin{align}
W \log\left(1 + \sum_{i \in \mathcal{C}} \rho |g_{c, k, i}|^2 + \sum_{(c', k') \in \mathcal{S}} \rho |h_{c, k, c', k'}|^2  \right) < CR.
\label{second_fcoc}
\end{align}



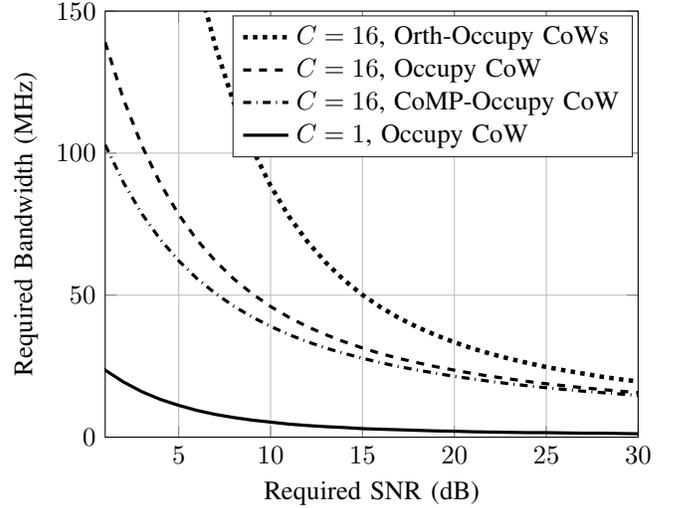
\begin{figure}[t]
\center

\begin{tikzpicture}

\begin{axis}[%
scale = 0.8,
width=\columnwidth,
height=0.8\columnwidth,
at={(0in,0in)},
scale only axis,
xmin=1,
xmax=30,
xlabel={Required SNR (dB)},
xmajorgrids,
ymin=0,
ymax=150,
ylabel={Required Bandwidth (MHz)},
ymajorgrids,
axis background/.style={fill=white},
legend style={at={(0.99,0.99)},anchor=north east,legend cell align=left,align=left,draw=white!15!black}
]
\addplot [color=black,dotted,line width=1.7pt]
  table[row sep=crcr]{%
1.0000  413.4000\\
    2.0000  338.4000\\
    3.0000  278.6000\\
    4.0000  230.9000\\
    5.0000  192.7000\\
    6.0000  162.1000\\
    7.0000  137.5000\\
    8.0000  117.7000\\
    9.0000  101.6000\\
   10.0000   88.5000\\
   11.0000   77.7000\\
   12.0000   68.9000\\
   13.0000   61.5000\\
   14.0000   55.3000\\
   15.0000   50.1000\\
   16.0000   45.7000\\
   17.0000   41.9000\\
   18.0000   38.7000\\
   19.0000   35.9000\\
   20.0000   33.4000\\
   21.0000   31.2000\\
   22.0000   29.3000\\
   23.0000   27.6000\\
   24.0000   26.1000\\
   25.0000   24.7000\\
   26.0000   23.5000\\
   27.0000   22.4000\\
   28.0000   21.3000\\
   29.0000   20.4000\\
   30.0000   19.6000\\
};
\addlegendentry{$C = 16$, Orth-Occupy CoWs};

\addplot [color=black, dashed, line width=1.2pt]
  table[row sep=crcr]{%
1.0000  139.0000\\
    2.0000  118.9000\\
    3.0000  102.6000\\
    4.0000   89.3000\\
    5.0000   78.4000\\
    6.0000   69.4000\\
    7.0000   62.0000\\
    8.0000   55.7000\\
    9.0000   50.5000\\
   10.0000   46.0000\\
   11.0000   42.2000\\
   12.0000   38.9000\\
   13.0000   36.1000\\
   14.0000   33.6000\\
   15.0000   31.4000\\
   16.0000   29.4000\\
   17.0000   27.7000\\
   18.0000   26.2000\\
   19.0000   24.8000\\
   20.0000   23.6000\\
   21.0000   22.4000\\
   22.0000   21.4000\\
   23.0000   20.5000\\
   24.0000   19.6000\\
   25.0000   18.8000\\
   26.0000   18.1000\\
   27.0000   17.4000\\
   28.0000   16.8000\\
   29.0000   16.2000\\
   30.0000   15.7000\\
};
\addlegendentry{$C = 16$, Occupy CoW};

\addplot [color=black, dashdotted,line width=1.2pt]
  table[row sep=crcr]{%
1.0000  102.9000\\
    2.0000   89.6000\\
    3.0000   78.6000\\
    4.0000   69.6000\\
    5.0000   62.1000\\
    6.0000   55.8000\\
    7.0000   50.6000\\
    8.0000   46.1000\\
    9.0000   42.3000\\
   10.0000   39.0000\\
   11.0000   36.1000\\
   12.0000   33.6000\\
   13.0000   31.4000\\
   14.0000   29.5000\\
   15.0000   27.8000\\
   16.0000   26.2000\\
   17.0000   24.8000\\
   18.0000   23.6000\\
   19.0000   22.5000\\
   20.0000   21.4000\\
   21.0000   20.5000\\
   22.0000   19.6000\\
   23.0000   18.8000\\
   24.0000   18.1000\\
   25.0000   17.4000\\
   26.0000   16.8000\\
   27.0000   16.2000\\
   28.0000   15.7000\\
   29.0000   15.2000\\
   30.0000   14.7000\\
};
\addlegendentry{$C = 16$, CoMP-Occupy CoW};

\addplot [color=black,solid,line width=1.2pt]
  table[row sep=crcr]{%
0   28.9000\\
    1.0000   23.6000\\
    2.0000   19.4000\\
    3.0000   16.0000\\
    4.0000   13.3000\\
    5.0000   11.2000\\
    6.0000    9.4000\\
    7.0000    8.0000\\
    8.0000    6.9000\\
    9.0000    6.0000\\
   10.0000    5.3000\\
   11.0000    4.6000\\
   12.0000    4.1000\\
   13.0000    3.7000\\
   14.0000    3.4000\\
   15.0000    3.0000\\
   16.0000    2.8000\\
   17.0000    2.6000\\
   18.0000    2.4000\\
   19.0000    2.2000\\
   20.0000    2.1000\\
   21.0000    1.9000\\
   22.0000    1.8000\\
   23.0000    1.7000\\
   24.0000    1.6000\\
   25.0000    1.6000\\
   26.0000    1.5000\\
   27.0000    1.4000\\
   28.0000    1.4000\\
   29.0000    1.3000\\
   30.0000    1.2000\\
};
\addlegendentry{$C =  1$, Occupy CoW};

\end{axis}

\end{tikzpicture}

\caption{Tradeoff between bandwidth and SNR for achieving $P_F = 10^{-9}$ by the interference avoidance benchmark protocols in a network with $C = 16$ cells, and in a single-cell scenario. Here, $K = 30$, $b = 160$bits, and $T = 1$ms.}
\label{fig_bwvspformula}
\end{figure}

\section{Bandwidth-Power Tradeoff of the Benchmark Protocols}
\label{sec_analysis}

In this section, we illustrate the tradeoff between the required bandwidth and the required SNR for achieving ultrareliability by the benchmark protocols in a multicell network. Characterizing very low failure probabilities of ultrareliable communication protocols can be a significant simulation burden. Assuming i.i.d. wireless channel gains throughout the network, a closed-form characterization of failure probability of the Occupy CoW protocol is provided in \cite{oc17}. This characterization does not consider the distance-dependent path loss of the wireless channel. Nevertheless, the closed-form formula captures important properties of the protocol. We first extend the failure probability formula of \cite{oc17} for the three benchmark protocols of the last section. 

Assume that channel gains $G_{c, k, i}$'s and $H_{c, k, c', k'}$'s are independent and are all equal in distribution to some random variable $H$. The assumption of identical distribution for channel gains can be valid when the nodes are in range of each other, in which case the channels may have a line-of-sight (LOS) component. If $H$ follows a Rician distribution, a realization of $H$ can be written as
\begin{align}
h = \sqrt{\frac{\kappa}{\kappa + 1}} e^{j {\theta}} + \sqrt{\frac{1}{\kappa + 1}} \bar{h},
\end{align}
where $\theta$ is drawn uniformly from $(0, 2\pi]$, $\bar{h}$ is drawn from $\mathcal{CN} (0, 1)$ distribution, and $\kappa$ is the Rician K-factor. The probability of outage for a link with spectral efficiency $CR/W$ (b/s/Hz) is
\begin{multline}
P_l = P\left( \frac{W}{C} \log\left(1 + \rho|h|^2\right) < R \right) \\ = 1-Q_1(\sqrt{\kappa}, \sqrt{2(\kappa + 1) R}),
\label{Pl}
\end{multline}
where $Q_1$ is the Marcum Q-function. 


In the Orth-Occupy CoWs scheme of Section \ref{sec_ooc}, the number of successful actuators of cell $c$ in the first phase ${A}_c \triangleq |\mathcal{S}_c^{orth}|$ are i.i.d. across $c$ with a binomial distribution. The probability of success for cell $c$ is \cite{oc17}:
\begin{align}
P_{S, 1} &= \sum_{k = 0}^{K} P({A}_c = k) P\left(\mathrm{``success"} \mid {A}_c = k\right)  \nonumber \\ &=  \sum_{k = 0}^{K} \binom{K}{k} \left(1 - P_l \right)^k P_{l}^{(K-k)} \left(1 - P_{l}^{k}\right)^{K-k},
\label{Ps_1}
\end{align}
and the probability of failure is
\begin{align}
P_{F, 1} = 1 - P_{S, 1}^C.
\label{Pf_1}
\end{align}
Here, obtaining spatial diversity by using distributed space-time coding is modeled by raising the probability of outage of a link to the power of the number of transmitters. 

In the Occupy CoW scheme of Section \ref{sec_oc}, the number of actuators that decoded the message of the $c$th controller in the first phase and are inside the cell ${A}_{c, i} \triangleq |\mathcal{S}_c^{orth}|$ and those that are outside the cell ${A}_{c, o} \triangleq |\mathcal{S}_c^{coop}|$ are each i.i.d. across $c$ with a binomial distribution. The probability of success for cell $c$ is
\begin{align}
P_{S, 2} &= \sum_{{k_o} = 0}^{CK-K} \sum_{{k_i} = 0}^{K}  P({{A}_{c, i}} = k_i) P({{A}_{c, o}} = k_o) \nonumber \\ & \ \ \ \ \ \ \ \ \ \ \ \ \ \ \ \ \ \ \ \ P\left(\mathrm{``success"} \mid {A}_{c, i} = k_i, {A}_{c, o} = k_o\right)  \nonumber \\ &=  \sum_{{k_o} = 0}^{CK-K} \sum_{{k_i} = 0}^{K} \binom{K}{k_i} \binom{CK-K}{k_o} \nonumber \\ & \left(1 - P_l \right)^{(k_i + k_o)} P_{l}^{(CK-k_i - k_o)} \left(1 - P_{l}^{(k_i + k_o)}\right)^{K-k_i},
\label{Ps_2}
\end{align}
and using the union bound, the failure probability is bounded by
\begin{align}
P_{F, 2} \leq \bar{P}_{F, 2} = C \left(1 - P_{S, 2}\right).
\label{Pf_2}
\end{align}

In CoMP-Occupy CoW scheme of Section \ref{sec_foc}, the number of successful actuators in the first phase ${A} \triangleq |\mathcal{S}|$ follows a binomial distribution. The probability of success is
\begin{align}
P_{S, 3} &= \sum_{k = 0}^{CK} P(A=k) P\left(\mathrm{``success"} \mid {A} = k\right)  \nonumber \\ &=  \sum_{k = 0}^{CK} \binom{CK}{k} \left(1 - P_l^C \right)^k P_{l}^{C( CK-k)} \left(1 - P_{l}^{k}\right)^{CK-k}.
\label{Ps_3}
\end{align}
The probability of failure of this protocol is
\begin{align}
P_{F, 3} = 1 - P_{S, 3}.
\label{Pf_3}
\end{align}

Fig.~\ref{fig_bwvspformula} shows the tradeoff between the required bandwidth and the required SNR for ultrareliable communication by the three interference avoidance benchmark schemes. For plotting this curve, we select the target probability of failure to be $P_F = 10^{-9}$, the number of actuators per cell to be $K = 30$, $b = 160$ bits, and $T = 1$ms, as in the scenario discussed in \cite{oc17}. Also, we select the K-factor in formula (\ref{Pl}) to be 4.7dB, based on A1-Indoor scenario of the WINNER II channel models \cite{winner}. The main observation from this figure is the significant increase in the required bandwidth for achieving ultrareliability at low SNRs when $C$ increases from 1 to 16. Since in these schemes inter-cell interference is avoided by orthogonalization, the required bandwidth increases linearly in the number of cells. Also, we observe that CoMP-Occupy CoW and Occupy CoW considerably outperform Orth-Occupy CoWs, indicating that inter-cell cooperation among actuators reduces the required bandwidth considerably.

\section{Successive Interference Cancellation for Ultrareliable Communication}

In all three protocols of Section \ref{sec_avoidance}, inter-cell interference is avoided by orthogonalization in the frequency domain. This orthogonalization results in a linear scaling of the required bandwidth for reliability with the number of cells, $C$. To alleviate the dependency of the bandwidth-SNR tradeoff for reliability on the number of cells, in this section we propose reusing full frequency in the network, then successively decoding and canceling as much inter-cell interference as possible at the receivers.

\subsection{Successive Interference Cancellation in Both Phases}
\label{sec_sic}

First, we consider full frequency reuse and successive interference cancellation in both the broadcast and cooperation phases. Since interference cancellation is used in both phases of the protocol, we refer to it as the IC-IC protocol. In the ideal scenario where the receivers could decode and cancel all of the interference, the bandwidth-power tradeoff for ultrareliability would not depend on the number of cells.




The protocol of this section is similar to the frequency division scheme of Section \ref{sec_ooc} in that both schemes use an Occupy CoW protocol in each cell. In the first phase, every controller concatenates all of its $K$ messages into a $Kb$-bit message and broadcasts this concatenated message to its $K$ actuators. In the second phase, those nodes that have decoded the message in the previous phase together with the controller retransmit the message for the rest of the receivers in that cell using distributed space-time coding. Each phase takes place within $T/2$ seconds of time and the aggregate rate in a cell is $R$ as in (\ref{rate}). The difference, here, is that all nodes transmit over the entire frequency band $W$ and interfere with each other, and we use successive interference cancellation in both phases. 


The successive interference cancellation process of the first phase is summarized in the first part of Algorithm \ref{SIC_Algo}. Here, we describe the decoding process for actuator $(c, k)$. During an iteration of the process, we denote the set of controllers for whom actuator $(c, k)$ has not decoded their messages by $\mathcal{I}_{(c, k)}$. Therefore, during the first iteration we have $\mathcal{I}_{(c, k)} \leftarrow \mathcal{C}$. In an iteration of the process, let $c^*$ be the index of the controller with highest signal-to-interference-plus-noise ratio (SINR), i.e.,
\begin{align}
c^*  = \arg\!\max\limits_{c' \in \mathcal{I}_{(c, k)}} R_{c, k, c'},
\label{c*}
\end{align}
where
\begin{align}
R_{c, k, c'} = W \log\left(1 + \frac{ \rho \left|g_{c, k, c'}\right|^2}{\sum\limits_{j \in  \mathcal{I}_{(c, k)}/\{c'\}} \rho \left|g_{c, k, j}\right|^2 + 1}\right).
\label{R1}
\end{align}
If $\max\limits_{c' \in \mathcal{I}_{(c, k)}} R_{c, k, c'} \geq R$, actuator $(c, k)$ decodes the message of controller $c^*$, cancels its contribution from the received signal, assigns $\mathcal{I}_{(c, k)} \leftarrow \mathcal{I}_{(c, k)} /\{c^*\}$, and proceeds to the next iteration. The process terminates for actuator $(c, k)$, when\footnote{Note that in the IC-IC protocol, the decoding process can be terminated for actuator $(c, k)$ once it decodes its intended message. But, for brevity, we use the first phase of the IC-IC protocol to describe the IC-IA protocol of the next section as well, in which an actuator decodes as many messages as it can.} $\max\limits_{c' \in \mathcal{I}_{(c, k)}} R_{c, k, c'} < R$. Finally, the set of successful actuators of cell $c$ that have decoded the $c$th controller's message in the first phase is
\begin{align}
\mathcal{S}_c^{icic} = \left\{(c,  k) \in \mathcal{K}: c \in \mathcal{C}/\mathcal{I}_{(c, k)} \right\}.
\label{first_icic}
\end{align}
where $\mathcal{C}/\mathcal{I}_{(c, k)}$ is the set of controllers whose messages are decoded at actuator $(c, k)$ by the end of the first phase.

The decoding process in the second phase is summarized in the second part of Algorithm \ref{SIC_Algo}. In this phase, only the unsuccessful receivers of the first phase continue to successively decode and cancel the interfering messages that they have not decoded in the previous phase until they decode their intended messages or they cannot decode any other message. For the unsuccessful receiver $(c, k)$, during an iteration of the process, we denote by $\mathcal{I}_{(c, k)}$ the set of controllers, and by $\mathcal{I}'_{(c, k)}$ the set of actuators whose messages have not been decoded by receiver $(c, k)$. Therefore, during the first iteration $\mathcal{I}_{(c, k)}$ is as in the last iteration of the previous phase and $\mathcal{I}'_{(c, k)} \leftarrow \cup_{i \in \mathcal{I}_{(c, k)} } \mathcal{S}_i^{icic}$. In an iteration of the process, let $c^*$ be the index of the message with the highest corresponding SINR for decoding, i.e.,
\begin{align}
c^*  = \arg\!\max\limits_{c' \in \mathcal{I}_{(c, k)}} R'_{c, k, c'},
\label{c*}
\end{align}
where
\begin{align}
&R'_{c, k, c'} = W \log \nonumber \\ & \left(1 + \frac{ \rho \left|g_{c, k, c'}\right|^2 + \sum\limits_{(i, j) \in \mathcal{S}_{c'}^{icic}} \rho \left|h_{c, k, i, j}\right|^2}{\sum\limits_{i \in  \mathcal{I}_{(c, k)}/\{c'\}} \rho \left|g_{c, k, i}\right|^2 + \sum\limits_{(i, j) \in  \mathcal{I}'_{(c, k)}/\mathcal{S}_{c'}^{icic}} \rho \left|h_{c, k, i, j}\right|^2 + 1}\right).
\label{R2}
\end{align}
Here, the impact of using distributed space-time coding for exploiting spatial diversity on both signal and interference is modeled by adding the received powers from each transmitter. If $\max\limits_{c' \in \mathcal{I}_{(c, k)}} R'_{c, k, c'} \geq R$, actuator $(c, k)$ decodes the message of controller $c^*$, subtracts its contribution from the received signal, assigns $\mathcal{I}_{(c, k)} \leftarrow \mathcal{I}_{(c, k)} /\{c^*\}$, $\mathcal{I}'_{(c, k)}  \leftarrow \mathcal{I}'_{(c, k)} / \mathcal{S}_{c^*}^{icic}$, and proceeds to the next iteration. The process terminates for actuator $(c, k)$ when it decodes its intended message or when $\max\limits_{c' \in \mathcal{I}_{(c, k)}} R'_{c, k, c'} < R$.

\begin{algorithm}[h]
\caption{Decoding process of the IC-IC protocol}
\label{SIC_Algo}
\textbf{1. Decoding in the first phase:}
\begin{algorithmic}[1]
\For{each cell $c$ in $\mathcal{C}$}

\State $\mathcal{S}_{c}^{icic}  \leftarrow \O$

\For{each actuator $k$ in cell $c$}

\State $\mathcal{I}_{(c, k)}  \leftarrow  \mathcal{C}$

\State $\mathcal{I}_{\mathrm{temp}}  \leftarrow  \O$

\While{$ \mathcal{I}_{(c, k)} \neq \mathcal{I}_{\mathrm{temp}}$}

\State $\mathcal{I}_{\mathrm{temp}}  \leftarrow \mathcal{I}_{(c, k)}$

\State $c^*  \leftarrow \arg\!\max\limits_{i \in \mathcal{I}_{(c, k)}} R_{c,k,i}$

\If{$R_{c,k,c^*} \geq R$}

\State $\mathcal{I}_{(c, k)}  \leftarrow \mathcal{I}_{(c, k)} /\{c^*\}$

\EndIf

\EndWhile

\If{$c \in \mathcal{C}/\mathcal{I}_{(c, k)}$}

\State $\mathcal{S}_{c}^{icic}  \leftarrow \mathcal{S}_{c}^{icic} \cup \{(c, k)\}$

\EndIf

\EndFor

\EndFor
\end{algorithmic}

\textbf{2. Decoding in the Second Phase:} 
\begin{algorithmic}[1]
\For{each cell $c$ in $\mathcal{C}$}

\For{each actuator $(c, k) \notin \mathcal{S}_{c}^{icic}$}

\State $\mathcal{I}'_{(c, k)}  \leftarrow  \cup_{i \in \mathcal{I}_{(c, k)} } \mathcal{S}_i^{icic}$

\State $\mathcal{I}_{\mathrm{temp}}  \leftarrow  \O$

\State $c^*  \leftarrow  0$

\While{$ \mathcal{I}'_{(c, k)} \neq \mathcal{I}_{\mathrm{temp}}$ \textbf{and} $c^* \neq c$}

\State $\mathcal{I}_{\mathrm{temp}}  \leftarrow \mathcal{I}'_{(c, k)}$

\State $c^*  \leftarrow \arg\!\max\limits_{c' \in \mathcal{I}_{(c, k)}} R'_{c, k, c'}$

\If{$ R'_{c, k, c^*} \geq R$}

\State $\mathcal{I}_{(c, k)}  \leftarrow \mathcal{I}_{(c, k)} /\{c^*\}$

\State $\mathcal{I}'_{(c, k)}  \leftarrow \mathcal{I}'_{(c, k)} / \mathcal{S}_{c^*}^{icic}$

\EndIf

\EndWhile

\EndFor

\EndFor

\end{algorithmic}
\end{algorithm}


\subsection{Interference Cancellation in the First Phase and Interference Avoidance in the Second Phase}
\label{sec_icia}

As we will see in the simulation results in Section \ref{sec_sim}, the IC-IC scheme of the previous section successfully alleviates the dependency of the bandwidth-SNR tradeoff for reliability on $C$, the number of cells in the network. However, since that receivers cannot decode some of the relatively weaker interference signals and treat them as noise, the failure probability of this scheme has a floor. Moreover, the diversity order of the IC-IC scheme in the second phase is limited by the number of successful nodes of the first phase in only one cell. To remove the saturation and improve the diversity order while having the improvement in the bandwidth-SNR tradeoff over the benchmark schemes, we propose a protocol with the interference cancellation scheme in its broadcast phase and the orthogonal frequency division interference avoidance scheme in its cooperation phase, and refer to it as the IC-IA protocol. In particular, we propose using the first phase of the IC-IC protocol of Section \ref{sec_sic} for the broadcast phase and using the second phase of the Occupy CoW protocol of Section \ref{sec_oc} for the cooperation phase.

Using the interference cancellation scheme in the broadcast phase of the protocol reduces the required transmit power for reliability. This is because for low transmit powers, the average number of decoded messages at a receiver in the first phase of the IC-IC scheme is larger than that of an inter-cell interference avoidance by frequency division scheme. In the IC-IA protocol, all the actuators that have decoded the message of the $c$th cell in the first phase relay this message over its associated frequency band in the second phase. The set of these actuators is
\begin{align}
\mathcal{S}_c^{icia} = \left\{(c',  k') \in \mathcal{K}: c \in \mathcal{C}/\mathcal{I}_{(c', k')} \right\},
\label{first_icia}
\end{align}
where $\mathcal{C}/\mathcal{I}_{(c', k')}$ is the set of controllers that their messages are decoded at actuator $(c', k')$ by the end of the first phase.

Using frequency division orthogonal transmission in the cooperation phase of this protocol improves the cooperative diversity order. This is because in the second phase of the Occupy CoW protocol an actuator relays all of the messages that it decoded in the first phase over their allocated frequency bands. In contrast, with full frequency reuse in the second phase, a half-duplex actuator can only relay one of its decoded messages. More importantly, interference avoidance by frequency division in the second phase removes the failure probability floor by providing the actuators with an interference-free version of their intended signal.

\section{Simulation Results}
\label{sec_sim}

In this section, we study the performance of the proposed interference management protocols by Monte Carlo simulations using a distance-dependent channel model for an industrial factory hall. Rather than simulating very low failure probabilities, we study the patterns of failure probability by varying transmit power spectral density $p$, number of cells $C$, and the bandwidth $W$.

\begin{figure*}[t]
\begin{tabular}{cc}

\begin{tikzpicture}

\begin{axis}[%
scale = 0.8,
width=0.8\columnwidth,
height=0.8\columnwidth,
at={(0in,0in)},
scale only axis,
xminorticks=false,
xtick={0, 10, 30, 40, 60, 70},
xticklabels={$0$, $10$, $D$, \ \ \ $D$+10 ,$2D$, \ \ \ \ \ $2D$+10},
x tick label style={anchor=north},
xmin=-10,
xmax=80,
xlabel={ meter\color{white}T B H p },
xmajorgrids,
yminorticks=false,
ytick={0, 10, 30, 40, 60, 70},
yticklabels={$0$, $10$, $D$, \ \ \ $D$+10 ,$2D$, \ \ \ \ \ $2D$+10},
y tick label style={anchor=east},
ymin=-10,
ymax=80,
xlabel={ meter },
ylabel={ meter },
ymajorgrids,
axis background/.style={fill=white},
legend style={at={(0.05,0.24)},anchor=south west,legend cell align=left,align=left,draw=white!15!black}
]

\addplot [color=blue,mark size=1.0pt,only marks,mark=*,mark options={solid,fill=white!49!black,draw=white!49!black}]
  table[row sep=crcr]{%
1.5	0.5\\
38.5	5.5\\
66.5	4.5\\
0.5	37.5\\
36.5	36.5\\
69.5	37.5\\
5.5	62.5\\
33.5	61.5\\
60.5	62.5\\
9.5	0.5\\
36.5	9.5\\
65.5	0.5\\
0.5	38.5\\
32.5	35.5\\
66.5	35.5\\
1.5	60.5\\
37.5	60.5\\
65.5	68.5\\
2.5	6.5\\
38.5	3.5\\
69.5	6.5\\
4.5	38.5\\
35.5	36.5\\
66.5	39.5\\
8.5	65.5\\
32.5	60.5\\
64.5	63.5\\
5.5	8.5\\
37.5	8.5\\
66.5	0.5\\
6.5	35.5\\
37.5	39.5\\
69.5	30.5\\
8.5	64.5\\
38.5	68.5\\
61.5	60.5\\
8.5	1.5\\
33.5	3.5\\
65.5	1.5\\
7.5	31.5\\
34.5	32.5\\
61.5	32.5\\
3.5	60.5\\
33.5	69.5\\
67.5	65.5\\
4.5	8.5\\
30.5	7.5\\
69.5	7.5\\
8.5	31.5\\
32.5	31.5\\
68.5	34.5\\
9.5	67.5\\
38.5	66.5\\
66.5	69.5\\
4.5	0.5\\
32.5	2.5\\
66.5	1.5\\
1.5	38.5\\
36.5	34.5\\
66.5	32.5\\
5.5	68.5\\
39.5	69.5\\
64.5	61.5\\
7.5	5.5\\
35.5	1.5\\
64.5	7.5\\
0.5	30.5\\
38.5	32.5\\
63.5	30.5\\
3.5	68.5\\
30.5	68.5\\
61.5	62.5\\
7.5	1.5\\
32.5	8.5\\
68.5	8.5\\
1.5	35.5\\
30.5	39.5\\
66.5	31.5\\
6.5	61.5\\
35.5	66.5\\
62.5	64.5\\
8.5	6.5\\
39.5	2.5\\
63.5	6.5\\
5.5	31.5\\
38.5	34.5\\
60.5	32.5\\
6.5	62.5\\
38.5	63.5\\
65.5	61.5\\
3.5	9.5\\
35.5	7.5\\
68.5	9.5\\
6.5	34.5\\
38.5	38.5\\
65.5	39.5\\
6.5	64.5\\
38.5	62.5\\
64.5	66.5\\
0.5	6.5\\
38.5	6.5\\
69.5	3.5\\
6.5	31.5\\
38.5	35.5\\
63.5	34.5\\
1.5	67.5\\
33.5	68.5\\
63.5	64.5\\
2.5	1.5\\
39.5	9.5\\
69.5	9.5\\
0.5	36.5\\
35.5	32.5\\
64.5	32.5\\
2.5	69.5\\
39.5	67.5\\
67.5	62.5\\
5.5	0.5\\
38.5	8.5\\
63.5	0.5\\
1.5	30.5\\
31.5	34.5\\
66.5	34.5\\
7.5	69.5\\
35.5	68.5\\
68.5	66.5\\
5.5	6.5\\
32.5	4.5\\
63.5	1.5\\
5.5	38.5\\
37.5	30.5\\
67.5	37.5\\
8.5	61.5\\
35.5	62.5\\
66.5	63.5\\
9.5	5.5\\
37.5	5.5\\
61.5	0.5\\
5.5	30.5\\
35.5	38.5\\
60.5	39.5\\
5.5	67.5\\
39.5	62.5\\
67.5	69.5\\
3.5	3.5\\
33.5	9.5\\
61.5	9.5\\
3.5	35.5\\
36.5	31.5\\
61.5	34.5\\
5.5	63.5\\
30.5	60.5\\
68.5	64.5\\
5.5	9.5\\
33.5	8.5\\
60.5	7.5\\
8.5	33.5\\
37.5	35.5\\
68.5	39.5\\
2.5	65.5\\
33.5	65.5\\
68.5	67.5\\
9.5	9.5\\
33.5	7.5\\
65.5	6.5\\
2.5	35.5\\
37.5	32.5\\
66.5	38.5\\
8.5	60.5\\
37.5	65.5\\
66.5	65.5\\
2.5	4.5\\
32.5	0.5\\
62.5	9.5\\
3.5	31.5\\
35.5	37.5\\
69.5	38.5\\
8.5	66.5\\
38.5	64.5\\
69.5	64.5\\
8.5	5.5\\
38.5	0.5\\
64.5	6.5\\
8.5	35.5\\
39.5	38.5\\
61.5	38.5\\
7.5	66.5\\
34.5	69.5\\
67.5	66.5\\
8.5	2.5\\
37.5	2.5\\
63.5	4.5\\
4.5	36.5\\
30.5	37.5\\
61.5	39.5\\
3.5	63.5\\
33.5	60.5\\
69.5	67.5\\
2.5	2.5\\
38.5	1.5\\
68.5	2.5\\
8.5	32.5\\
37.5	36.5\\
61.5	36.5\\
4.5	66.5\\
31.5	67.5\\
68.5	60.5\\
7.5	7.5\\
35.5	2.5\\
61.5	3.5\\
7.5	35.5\\
39.5	34.5\\
63.5	35.5\\
4.5	61.5\\
30.5	67.5\\
66.5	64.5\\
6.5	9.5\\
36.5	7.5\\
62.5	4.5\\
9.5	37.5\\
32.5	33.5\\
68.5	30.5\\
7.5	65.5\\
37.5	69.5\\
63.5	69.5\\
1.5	3.5\\
31.5	1.5\\
60.5	6.5\\
1.5	34.5\\
36.5	30.5\\
67.5	39.5\\
9.5	63.5\\
36.5	63.5\\
62.5	69.5\\
3.5	7.5\\
32.5	1.5\\
61.5	4.5\\
0.5	31.5\\
34.5	38.5\\
63.5	39.5\\
8.5	63.5\\
33.5	63.5\\
61.5	61.5\\
8.5	9.5\\
39.5	1.5\\
60.5	8.5\\
3.5	38.5\\
36.5	33.5\\
64.5	31.5\\
0.5	67.5\\
32.5	64.5\\
62.5	60.5\\
6.5	4.5\\
39.5	6.5\\
60.5	5.5\\
7.5	37.5\\
31.5	38.5\\
67.5	31.5\\
3.5	61.5\\
30.5	63.5\\
60.5	69.5\\
2.5	5.5\\
33.5	5.5\\
66.5	2.5\\
3.5	33.5\\
34.5	39.5\\
66.5	37.5\\
2.5	63.5\\
31.5	61.5\\
67.5	63.5\\
};
\addlegendentry{Actuators};

\addplot [color=black,mark size=2.5pt,only marks,mark=triangle*,mark options={solid,fill=black}]
  table[row sep=crcr]{%
5	5\\
35	5\\
65	5\\
5	35\\
35	35\\
65	35\\
5	65\\
35	65\\
65	65\\
};
\addlegendentry{Controllers};

\end{axis}
\end{tikzpicture}

 \ \ \ \  

\begin{tikzpicture}

\begin{axis}[%
scale = 0.8,
width=\columnwidth,
height=0.8\columnwidth,
at={(0in,0in)},
scale only axis,
xmin=10,
xmax=20,
xlabel={Average SNR of a link (dB)},
xmajorgrids,
ymode=log,
ymin=1e-07,
ymax=1,
yminorticks=true,
ylabel={Failure probability in the 9 cells},
ymajorgrids,
axis background/.style={fill=white},
legend style={at={(0.02,0.02)},anchor=south west,legend cell align=left,align=left,draw=white!15!black}
]
\addplot [color=black, solid, line width=1.2pt]
  table[row sep=crcr]{%
10	1\\
12	0.970873786407767\\
14	0.8\\
16	0.602409638554217\\
18	0.367647058823529\\
20	0.261096605744125\\
};
\addlegendentry{$D = 100$m};

\addplot [color=black, dashdotted,line width=2.0pt]
  table[row sep=crcr]{%
10	0.980392156862745\\
12	0.645161290322581\\
14	0.136239782016349\\
16	0.0206228088265622\\
18	0.00399392922757409\\
20	0.000601956358164033\\
};
\addlegendentry{$D = 250$m};

\addplot [color=black, dashed, line width=1.5pt]
  table[row sep=crcr]{%
10	0.970873786407767\\
12	0.480769230769231\\
14	0.0599880023995201\\
16	0.004293688278231\\
18	0.000162431271268345\\
20	6.75780467308951e-06\\
};
\addlegendentry{$D = 500$m};

\addplot [color=black, dotted, line width=1.7pt]
  table[row sep=crcr]{%
10	1\\
12	0.347222222222222\\
14	0.048828125\\
16	0.00203832042397065\\
18	4.28567632686682e-05\\
20	6.71074282554485e-07\\
};
\addlegendentry{No Interference};

\end{axis}
\end{tikzpicture}

\end{tabular}
\caption{Impact of treating inter-cell interference as noise on reliability. The network layout of 9 interfering cells with $K = 30$ users per cell is shown on the left and the simulation results of using Occupy CoW protocol in each cell on the right. Here, $W = 2$MHz, $b = 160$bits, and $T = 1$ms. In this example, achieving ultrareliability in presence of inter-cell interference that is treated as noise in both phases is possible only at cell distances as long as $D = 500$m.}
\label{fig_intisimpo}
\end{figure*}

\subsection{Channel Model}
\label{sec_chnmod}

To simulate fading in both controller-to-actuator and actuator-to-actuator wireless links, we use A1-Indoor scenario of the WINNER II channel models \cite{winner}. The existence of LOS component in this model follows a distance-dependent probability distribution. Let the distance between transmitter and receiver be $d$ m. Then, the probability of having a LOS component in this model is given by
\begin{align}
P_{LOS} = \begin{cases} 1  &  d \leq 2.5  \\ 1-0.9 \left( 1 - \left(1.24-0.61 \log_{10}(d)\right)^3 \right)^{\frac{1}{3}} & d > 2.5 \end{cases}
\label{Pr_LOS}
\end{align}
We set carrier frequency to $3$ GHz. When the LOS component does not exist, the path loss model is
\begin{align}
PL_{NLOS} = 36.8 \log_{10}(d) + 43.8 + 20 \log_{10}\left(0.6\right),
\label{PL_NLOS}
\end{align}
and the standard deviation of shadow fading is 4 dB. When the LOS component exists, the path loss model is
\begin{align}
PL_{LOS} = 18.7 \log_{10}(d) + 46.8 + 20 \log_{10}\left(0.6\right),
\label{PL_LOS}
\end{align}
the Ricean K-factor is 4.7dB. Even in this case, there is a shadow fading component with 3dB standard deviation. In both LOS and non-LOS cases, we have Rayleigh fading, which is i.i.d. across links. The power spectral density of background noise is $\sigma^2 = -169$dBm/Hz. In our simulations, the minimum distance between any two points is set to 1m.

We simulate a square-shaped factory hall with $C$ square-shaped cells. The side of each cell is 10m. Hence, the side of the factory hall is $10\sqrt{C}$m. The controller of each cell is located at the center of the square-shaped cell and its associated actuators are distributed uniformly in that cell. In this setting, we note through simulations that the average path loss between a transmitter-receiver pair is $\overline{PL} = 59$dB. Therefore, the average SNR of a link is given by
\begin{align}
\overline{SNR} = p - \overline{PL}  - \sigma^2 = p + 110 \ \ \  \mathrm{dB},
\label{SNR_avg}
\end{align}
where $p$ is the transmit power spectral density in dBm/Hz. Note that because of the very short distance between transmitter-receiver pairs, the transmit power required to achieve a typical value of SNR is fairly low, e.g., for achieving $\overline{SNR} = 5$dB, we need $p = -105$dBm/Hz, which corresponds to $0.95 \mu W$ of transmit power over $W = 30$MHz of bandwidth.

\subsection{Impact of Treating Inter-cell Interference as Noise}
\label{sec_tin}

In this section, we study the failure probability of 9 interfering cells with Occupy CoW protocol in each, assuming the inter-cell interference is treated as noise in both phases of the protocol. To model the reuse of frequency in distant cells within a geographical area, we simulate nine 10m-by-10m interfering cells in a $3D$-by-$3D$ square-shaped geographical area with wraparound. The parameter $D$ determines the distance between the cells; the location of a cell is obtained by either horizontal or vertical translation of a closest cell by $D$. Fig.~\ref{fig_intisimpo} shows the simulation layout and the numerical results. We observe that treating inter-cell interference as noise results in a failure probability floor, the value of which depends on the distance between cells. In our simulations, for approaching the interference-free performance the distance $D$ should be as long as 500m. Therefore, the conclusion here is that frequency reuse may not be useful for indoor applications. This observation motivates the design of interference mitigation schemes for ultrareliable low-latency wireless communication for factory automation. In the rest of the simulations, we consider full reuse of frequency throughout the network without any separation between adjacent cells. 

\begin{figure*}[t]
\begin{tabular}{c}

\begin{tikzpicture}

\begin{axis}[%
scale = 0.9,
width=2\columnwidth,
height=0.9\columnwidth,
at={(0in,0in)},
scale only axis,
xmin=-15,
xmax=50,
xlabel={Average SNR of a link (dB)},
xmajorgrids,
ymode=log,
ymin=1e-4,
ymax=1,
yminorticks=true,
ylabel={Probability of failure},
ymajorgrids,
axis background/.style={fill=white},
legend style={at={(0.01,0.01)},anchor=south west,legend cell align=left,align=left,draw=white!15!black, font=\fontsize{6}{5}\selectfont}
]

\addplot [color=black,solid,line width=1.0pt]
  table[row sep=crcr]{%
-17.000000000000000   1.000000000000000\\
 -16.000000000000000   0.998003992015968\\
 -15.000000000000000   0.990099009900990\\
 -14.000000000000000   0.967117988394584\\
 -13.000000000000000   0.889679715302491\\
 -12.000000000000000   0.688705234159780\\
 -11.000000000000000   0.448028673835125\\
 -10.000000000000000   0.255232261357836\\
  -9.000000000000000   0.108837614279495\\
  -8.000000000000000   0.039569484013928\\
  -7.000000000000000   0.012848185836160\\
  -6.000000000000000   0.003421727972626\\
  -5.000000000000000   0.000689117185756\\
  -3.000000000000000   0.000019600425721\\
};
\addlegendentry{$C=1$, Occupy CoW};

\addplot [color=red, dashed, line width=1.0pt,mark size=2pt,mark=triangle,mark options={solid,rotate=180}]
  table[row sep=crcr]{%
0   0.996015936254980\\
   1.000000000000000   0.815660685154976\\
   2.000000000000000   0.304321363359708\\
   3.000000000000000   0.043671936413661\\
   4.000000000000000   0.003100294527980\\
   4.500000000000000   0.000601793344166\\
   4.900000000000006   0.000098325516455\\
};
\addlegendentry{$C = 9$, CoMP-Occupy CoW};

\addplot [color=black,solid,line width=1.0pt,mark size=1.7pt,mark=triangle,mark options={solid,rotate=180}]
  table[row sep=crcr]{%
-10.000000000000000   1.000000000000000\\
  -9.000000000000000   0.934579439252336\\
  -8.000000000000000   0.745156482861401\\
  -7.000000000000000   0.498504486540379\\
  -6.000000000000000   0.266666666666667\\
  -5.000000000000000   0.125000000000000\\
  -3.000000000000000   0.015304560759106\\
                   0   0.000886627004885\\
   3.500000000000000   0.000057390785336\\
};
\addlegendentry{$C = 9$, IC-IC};

\addplot [color=green, dashdotted, line width=1.0pt,mark size=2pt,mark=triangle,mark options={solid,rotate=180}]
  table[row sep=crcr]{%
-6.000000000000000   1.000000000000000\\
  -4.000000000000000   0.996015936254980\\
  -2.000000000000000   0.563063063063063\\
                   0   0.071042909917590\\
   2.000000000000000   0.003051757812500\\
   3.000000000000000   0.000578328860923\\
};
\addlegendentry{$C = 9$, IC-IA};

\addplot [color=blue, dotted, line width=1.0pt,mark size=2pt,mark=triangle,mark options={solid,rotate=180}]
  table[row sep=crcr]{%
3.000000000000000   0.990099009900990\\
   4.000000000000000   0.846023688663283\\
   5.000000000000000   0.474833808167141\\
   7.000000000000000   0.036403349108118\\
   9.000000000000000   0.000950977129000\\
  10.000000000000000   0.000120752529765\\
};
\addlegendentry{$C = 9$, Occupy CoW};

\addplot [color=black,solid,line width=1.0pt, mark=o ,mark options={solid}]
  table[row sep=crcr]{%
-9.000000000000000   1.000000000000000\\
  -8.000000000000000   0.980392156862745\\
  -7.000000000000000   0.917431192660551\\
  -6.000000000000000   0.724637681159420\\
  -5.000000000000000   0.513347022587269\\
  -3.000000000000000   0.214592274678112\\
                   0   0.041928721174004\\
   5.000000000000000   0.003640202395253\\
  10.000000000000000   0.000446690304029\\
  15.000000000000000   0.000100928759328\\
};
\addlegendentry{$C = 16$, IC-IC};

\addplot [color=red, dashed, line width=1.0pt,mark size=2pt,mark=triangle,mark options={solid}]
  table[row sep=crcr]{%
20.000000000000000   0.000008939290596\\
  19.500000000000000   0.000214114422748\\
  19.000000000000000   0.003192338387869\\
  18.000000000000000   0.080321285140562\\
  17.000000000000000   0.571428571428571\\
  16.500000000000000   0.909090909090909\\
  16.000000000000000   1.000000000000000\\
};
\addlegendentry{$C = 25$, CoMP-Occupy CoW};

\addplot [color=black,solid,line width=1.0pt,mark size=1.7pt,mark=triangle,mark options={solid}]
  table[row sep=crcr]{%
-6.000000000000000   0.980392156862745\\
  -5.000000000000000   0.892857142857143\\
  -3.000000000000000   0.581395348837209\\
                   0   0.224215246636771\\
   5.000000000000000   0.032701111837802\\
  10.000000000000000   0.004716313729189\\
  15.000000000000000   0.001438497058274\\
  20.000000000000000   0.000573374569969\\
30   			        0.00035\\
  40.000000000000000   0.000264823495140\\
};
\addlegendentry{$C = 25$, IC-IC};

\addplot [color= green, dashdotted, line width=1.0pt,mark size=2pt,mark=triangle,mark options={solid}]
  table[row sep=crcr]{%
7.000000000000000   1.000000000000000\\
   8.000000000000000   0.927643784786642\\
   9.000000000000000   0.615763546798030\\
  10.000000000000000   0.205592105263158\\
  12.000000000000000   0.018180495963930\\
  14.000000000000000   0.001126633618747\\
  16.000000000000000   0.000035670589242\\
};
\addlegendentry{$C = 25$, IC-IA};

\addplot [color= blue, dotted, line width=1.0pt,mark size=2pt,mark=triangle,mark options={solid}]
  table[row sep=crcr]{%
20.000000000000000   0.990099009900990\\
  21.000000000000000   0.862068965517241\\
  22.000000000000000   0.331125827814570\\
  23.000000000000000   0.076628352490421\\
  24.000000000000000   0.013503375843961\\
26   0.000134\\ 
};
\addlegendentry{$C = 25$, Occupy CoW};

\addplot [color=black,solid,line width=1.0pt,mark=asterisk,mark options={solid}]
  table[row sep=crcr]{%
-5.000000000000000   0.990654205607477\\
  -3.000000000000000   0.847457627118644\\
                   0   0.527426160337553\\
   5.000000000000000   0.143307537976498\\
  10.000000000000000   0.034705351565211\\
  15.000000000000000   0.010271158586689\\
  20.000000000000000   0.005120852109791\\
  25.000000000000000   0.004012776680952\\
  30.000000000000000   0.003501792917974\\
  45.000000000000000   0.002940484591861\\
};
\addlegendentry{$C = 36$, IC-IC};

\addplot [color=red, dashed, line width=1.0pt,mark size=2pt,mark=triangle,mark options={solid,rotate=270}]
  table[row sep=crcr]{%
39.000000000000000   1.000000000000000\\
  40.000000000000000   0.487804878048780\\
  41.000000000000000   0.035842293906810\\
  42.000000000000000   0.000245224257584\\
  42.099999999999994   0.000109586639197\\
  42.200000000000003   0.000060368610737\\
};
\addlegendentry{$C = 49$, CoMP-Occupy CoW};

\addplot [color=black,solid,line width=1.0pt,mark size=1.7pt,mark=triangle,mark options={solid,rotate=270}]
  table[row sep=crcr]{%
-3.000000000000000   0.980392156862745\\
                   0   0.821018062397373\\
   5.000000000000000   0.340136054421769\\
  10.000000000000000   0.128139415684264\\
  15.000000000000000   0.053688392569526\\
  20.000000000000000   0.032663316582915\\
  25.000000000000000   0.027348394768133\\
  30.000000000000000   0.024146279306830\\
  35.000000000000000   0.024651590849329\\
  40.000000000000000   0.022961416739613\\
  45.000000000000000   0.020973731261175\\
};
\addlegendentry{$C = 49$, IC-IC};

\addplot [color= green, dashdotted ,line width=1.0pt,mark size=2pt,mark=triangle,mark options={solid,rotate=270}]
  table[row sep=crcr]{%
28.000000000000000   0.978473581213307\\
  29.000000000000000   0.699300699300699\\
  30.000000000000000   0.339443312966735\\
  31.000000000000000   0.094876660341556\\
  36.000000000000000   0.000412379636694\\
  37.000000000000000   0.000148587306186\\
};
\addlegendentry{$C = 49$, IC-IA};

\addplot [color=blue, dotted, line width=1.0pt,mark size=2pt,mark=triangle,mark options={solid,rotate=270}]
  table[row sep=crcr]{%
43.000000000000000   1.000000000000000\\
  44.000000000000000   0.952380952380952\\
  45.000000000000000   0.615006150061501\\
  47.000000000000000   0.024014617593318\\
  49.000000000000000   0.000366434591425\\
};
\addlegendentry{$C = 49$, Occupy CoW};

\end{axis}
\end{tikzpicture}

\end{tabular}
\caption{Probability of failure versus average SNR of a link for Occupy CoW, CoMP-Occupy CoW, IC-IC, and IC-IA protocols for different number of cells $C$. Here, $W = 30$MHz, $K = 30$, $b = 160$ bits, $T = 1$ms.}
\label{fig_overc}
\end{figure*}
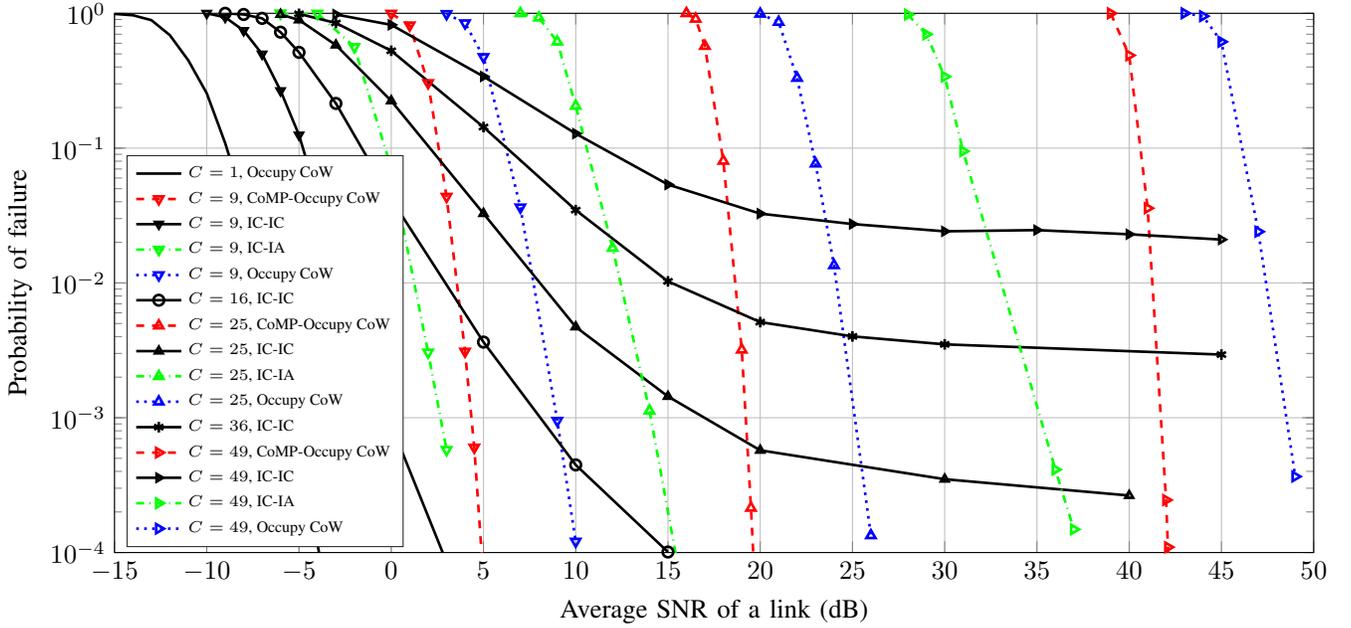

\subsection{Dependency of Performance on Number of Cells}
\label{sec_deponc}

Fig.~\ref{fig_overc} shows the probability of failure versus transmit power for CoMP-Occupy CoW, Occupy CoW, IC-IC, and IC-IA protocols for different number of cells $C$ at $W = 30$MHz, $K = 30$, $b = 160$bits, and $T = 1$ms.

In Fig.~\ref{fig_overc}, we observe that CoMP-Occupy CoW and Occupy CoW have the highest diversity order. However, at a given bandwidth, the required transmit power (in dB) for reliability of these two protocols increases linearly in number of cells $C$, for large values of $C$. The high dependency of bandwidth-power tradeoff in number of cells is the main drawback of these schemes. For this reason, they may not be the preferred protocols, despite having the largest diversity order. 


Here, we observe that the dependency of the required transmit power for reliability on the number of cells in the IC-IC protocol is the smallest, but still some dependency and a failure probability floor exist. In the ideal case, this dependency of the required transmit power on $C$ and the failure probability floor would not exist, if actuators could decode and cancel all of the interference. In reality, however, increasing $C$ increases the number of uncanceled interferers. This in turn reduces the SINR for decoding the intended message at receivers. More transmit power is needed to cancel the impact of this reduction in SINR and to obtain low failure probabilities. However, further increasing the power of transmitters results in a failure probability floor. This scheme has the best bandwidth-power tradeoff in the low-power regime.

The IC-IA protocol shows  improved scalability of transmit power as compared to the interference avoidance protocols. This is due to full bandwidth reuse and successive interference cancellation in the first phase. Orthogonal frequency division in the second phase, not only removes the failure probability floor of the IC-IC protocol by providing actuators with an interference-free version of their intended signal, but also improves the diversity order of the IC-IC protocol by enabling half-duplex actuators to relay all of the messages that they have decoded. Therefore, it has the main virtue of the Occupy CoW schemes, i.e., high diversity order without a failure probability floor, and the main virtue of the interference cancellation scheme, i.e., working at lower transmit powers, simultaneously. This scheme has the best bandwidth-power tradeoff at the medium-power regime.  



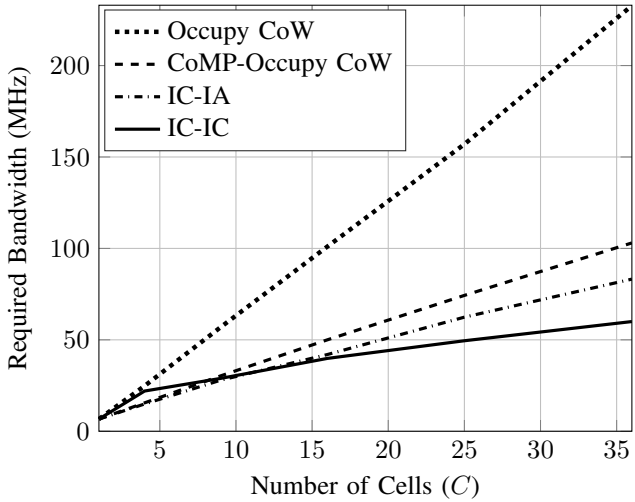
\begin{figure}[t]
\center

\begin{tikzpicture}

\begin{axis}[%
scale = 0.8,
width=\columnwidth,
height=0.8\columnwidth,
at={(0in,0in)},
scale only axis,
xmin=1,
xmax=36,
xlabel={Number of Cells ($C$)},
xmajorgrids,
ymin=0,
ymax=233,
ylabel={Required Bandwidth (MHz)},
ymajorgrids,
axis background/.style={fill=white},
legend style={at={(0.01,0.99)},anchor=north west,legend cell align=left,align=left,draw=white!15!black}
]
\addplot [color=black,dotted,line width=1.7pt]
  table[row sep=crcr]{%
1	6.67\\
4	24.8\\
9	56.96\\
16	101\\
25	157\\
36	232.75\\
};
\addlegendentry{Occupy CoW};

\addplot [color=black, dashed, line width=1.2pt]
  table[row sep=crcr]{%
1	6.67\\
4	15.54\\
9	30.31\\
16	50\\
25	74.3\\
36	103\\
};
\addlegendentry{CoMP-Occupy CoW};

\addplot [color=black, dashdotted,line width=1.2pt]
  table[row sep=crcr]{%
1	6.67\\
4	14.92\\
9	27.94\\
16	41.96\\
25	62.4\\
36	83.2\\
};
\addlegendentry{IC-IA};

\addplot [color=black,solid,line width=1.2pt]
  table[row sep=crcr]{%
1	6.67\\
4	22\\
9	28.92\\
16	39.84\\
25	49.5\\
36	60\\
};
\addlegendentry{IC-IC};

\end{axis}

\end{tikzpicture}

\caption{The required bandwidth for having $P_F = 5 \times 10^{-5}$ versus the number of cells $C$. Here, $\overline{SNR} = 5$dB, $K = 30$, $b = 160$bits, and $T = 1$ms.}
\label{fig_bwvsC}
\end{figure}

Fig.~\ref{fig_bwvsC} shows the scaling of the required bandwidth with number of cells $C$ for achieving $P_F = 5 \times 10^{-5}$ at average SNR of a link $\overline{SNR} = 5$ dB. Here, $P_F = 5 \times 10^{-5}$ is the smallest failure probability for which we can run Mont Carlo simulations in a reasonable time. For the Occupy CoW protocol this required bandwidth has almost a linear scaling in $C$ with a large slope. The slopes of the other curves are smaller. We observe a notable saving in the required bandwidth for achieving ultrareliability by the two proposed protocols over the two benchmark protocols.

\subsection{Dependency of Performance on Bandwidth}
\label{sec_deponw}

Fig.~\ref{fig_overw16} shows the probability of failure of CoMP-Occupy CoW, IC-IC, and IC-IA protocols versus average SNR for different values of bandwidth $W$, $K = 30$, $b = 160$bits, and $T = 1$ms, with $C = 16$ cells.

We observe from Fig.~\ref{fig_overw16} that by increasing bandwidth the floor of failure probability of IC-IC protocol decreases. Also, increasing bandwidth increases the diversity order of the IC-IA protocol. These are due to the fact that at larger bandwidths, more inter-cell interference can be decoded.

Increasing bandwidth reduces the required transmit power for achieving ultrareliability by CoMP-Occupy CoW and IC-IA protocols. However, at larger bandwidths, this reduction in the transmit power becomes slower. As consequence, there are significant regions in which IC-IC outperforms both IC-IA and CoMP-Occupy CoW.

Increasing bandwidth reduces the required transmit power for achieving ultrareliability. For CoMP-Occupy CoW and IC-IA protocols, however, this reduction in the transmit power becomes much slower for larger bandwidths. As a consequence, there are significant regions in which IC-IC outperforms both IC-IA and CoMP-Occupy CoW.

\begin{figure*}[t]
\begin{tabular}{c}

\begin{tikzpicture}

\begin{axis}[%
scale = 0.9,
width=2\columnwidth,
height=0.9\columnwidth,
at={(0in,0in)},
scale only axis,
scale only axis,
xmin=-10,
xmax=30,
xtick distance=5,
xlabel={Average SNR of a link (dB)},
xmajorgrids,
ymode=log,
ymin=1e-03,
ymax=1,
yminorticks=true,
ylabel={Probability of failure},
ymajorgrids,
axis background/.style={fill=white},
legend style={at={(0.01,0.01)},anchor=south west,legend cell align=left,align=left,draw=white!15!black, font=\fontsize{6}{5}\selectfont}
]

\addplot [color=black,solid,line width=1.0pt,mark size=1pt,mark=o,mark options={solid}]
  table[row sep=crcr]{%
-10	1\\
-5	0.505050505050505\\
0	0.0387897595034911\\
5	0.00392441860465116\\
10	0.000364890259254529\\
};
\addlegendentry{30MHz,IC-IC};

\addplot [color=red,dashed,line width=1.0pt,mark size=1pt,mark=o,mark options={solid}]
  table[row sep=crcr]{%
7.5	0.968992248062015\\
8	0.866551126516464\\
8.5	0.54585152838428\\
9	0.264690312334569\\
9.5	0.11001100110011\\
10	0.0249438762783737\\
11	0.00047631522541618\\
11.3  0.0001611\\
};
\addlegendentry{30MHz, CoMP-Occupy CoW};

\addplot [color=green,dashdotted,line width=1.0pt,mark size=1pt,mark=o,mark options={solid}]
  table[row sep=crcr]{%
0	1\\
1	1\\
2	0.87260034904014\\
3	0.535331905781585\\
4	0.178062678062678\\
5	0.049563838223632\\
7	0.00323991576219018\\
9	3.96375542043554e-05\\
};
\addlegendentry{30MHz, IC-IA};

\addplot [color=black,solid,line width=1.0pt,mark=triangle,mark options={solid,rotate=180}]
  table[row sep=crcr]{%
-10	1\\
-5	0.6765899864682\\
0	0.0837801608579088\\
5	0.0086773919230836\\
10	0.00108037464604661\\
15	0.000213587587998086\\
};
\addlegendentry{28MHz, IC-IC};

\addplot [color=black,solid,line width=1.0pt,mark=star,mark options={solid}]
  table[row sep=crcr]{%
-10	1\\
-5	0.871080139372822\\
0	0.164365548980934\\
5	0.021323780279768\\
10	0.00330745109750635\\
15	0.00073613309286319\\
};
\addlegendentry{26MHz,IC-IC};

\addplot [color=red, dashed, line width=1.0pt,mark=star,mark options={solid}]
  table[row sep=crcr]{%
10	1\\
10.8	0.576923076923077\\
11.6	0.173410404624277\\
12.4	0.0229007633587786\\
13.2	0.00146756677428823\\
14	6e-05\\
};
\addlegendentry{26MHz, CoMP-Occupy CoW};

\addplot [color= green, dashdotted, line width=1.0pt,mark=star,mark options={solid}]
  table[row sep=crcr]{%
2	1\\
4	0.841750841750842\\
5	0.479386385426654\\
6	0.162654521795706\\
8	0.00766871165644172\\
10	0.000370388661168453\\
};
\addlegendentry{26MHz, IC-IA};

\addplot [color=black,solid,line width=1.0pt,mark=triangle,mark options={solid}]
  table[row sep=crcr]{%
-10	1\\
-5	0.957854406130268\\
0	0.32133676092545\\
5	0.0611172228333944\\
10	0.0107902801156718\\
15	0.00290340281588398\\
20	0.00128261506889475\\
};
\addlegendentry{24MHz, IC-IC};

\addplot [color=black,solid,line width=1.0pt,mark=o,mark options={solid}]
  table[row sep=crcr]{%
-10	1\\
-5	0.99009900990099\\
0	0.591016548463357\\
5	0.174764068507515\\
10	0.045028818443804\\
15	0.0148038490007402\\
20	0.00770986245605378\\
25	0.00555370432078196\\
35     0.0051\\
};
\addlegendentry{22MHz, IC-IC};

\addplot [color=red, dashed, line width=1.0pt,mark=o,mark options={solid}]
  table[row sep=crcr]{%
12.8	1\\
13.8	0.882352941176471\\
14.8	0.20979020979021\\
15.8	0.0201477501678979\\
16.8	0.000472604681937049\\
};
\addlegendentry{22MHz, CoMP-Occupy CoW};

\addplot [color=green, dashdotted,line width=1.0pt,mark=o,mark options={solid}]
  table[row sep=crcr]{%
5	1\\
6	0.963391136801541\\
7	0.702247191011236\\
8	0.323834196891192\\
9	0.109529025191676\\
10	0.0338983050847458\\
12	0.0013447487561074\\
};
\addlegendentry{22MHz, IC-IA};

\addplot [color=black,solid,line width=1.0pt,mark=triangle,mark options={solid,rotate=270}]
  table[row sep=crcr]{%
-10	1\\
-5	1\\
0	0.880281690140845\\
5	0.431034482758621\\
10	0.16463615409944\\
15	0.0755857898715042\\
20	0.0473933649289099\\
25	0.0400288207509407\\
30	0.034952813701503\\
35	0.036613942589338\\
40	0.0357500357500357\\
};
\addlegendentry{20MHz, IC-IC};

\addplot [color=black,solid,line width=1.0pt,mark size=1.5pt,mark=square,mark options={solid}]
  table[row sep=crcr]{%
-10	1\\
-15	1\\
0	0.978473581213307\\
5	0.827814569536424\\
10	0.501002004008016\\
15	0.278396436525612\\
20	0.209205020920502\\
25	0.182149362477231\\
30	0.189753320683112\\
35	0.186081131373279\\
40	0.174520069808028\\
};
\addlegendentry{18MHz, IC-IC};

\addplot [color=red, dashed,line width=1.0pt,mark size=1.5pt,mark=square,mark options={solid}]
  table[row sep=crcr]{%
17	1\\
18	0.909090909090909\\
19	0.465116279069767\\
20	0.0651465798045603\\
21	0.00314169022934339\\
21.5  0.0006435\\
};
\addlegendentry{18MHz, CoMP-Occupy CoW};

\addplot [color= green, dashdotted,line width=1.0pt,mark size=1.5pt,mark=square,mark options={solid}]
  table[row sep=crcr]{%
9	1\\
10	0.94876660341556\\
11	0.677506775067751\\
12	0.301932367149758\\
14	0.0281425891181989\\
16	0.00181028240405503\\
};
\addlegendentry{18MHz, IC-IA};

\addplot [color=black,solid,line width=1.5pt]
  table[row sep=crcr]{%
-10	1\\
-5	1\\
0	1\\
5	1\\
10	0.909090909090909\\
15	0.726744186046512\\
20	0.632111251580278\\
25	0.608272506082725\\
30	0.586166471277843\\
35	0.580720092915215\\
40	0.566893424036281\\
};
\addlegendentry{16MHz, IC-IC};

\addplot [color=red, dashed,line width=1.0pt,mark size=3pt,mark=x,mark options={solid}]
  table[row sep=crcr]{%
25	1\\
25.5	0.952380952380952\\
26	0.571428571428571\\
26.5	0.307692307692308\\
27	0.16\\
27.5	0.044543429844098\\
28	0.00890075656430797\\
29   0.0002497\\
};
\addlegendentry{14MHz, CoMP-Occupy CoW};

\addplot [color= green, dashdotted, line width=1.0pt,mark size=3pt,mark=x,mark options={solid}]
  table[row sep=crcr]{%
16	1\\
17	0.943396226415094\\
18	0.684931506849315\\
19	0.335345405767941\\
20	0.136948781155848\\
21	0.0398247710075667\\
23	0.00360750360750361\\
24	0.00120170642312083\\
};
\addlegendentry{14MHz, IC-IA};

\end{axis}
\end{tikzpicture}

\end{tabular}
\caption{Probability of failure versus average SNR of a link for CoMP-Occupy CoW, IC-IC, and IC-IA protocols for different bandwidths, with $C = 16$ cells, $K = 30$, $b = 160$bits, $T = 1$ms.}
\label{fig_overw16}
\end{figure*}
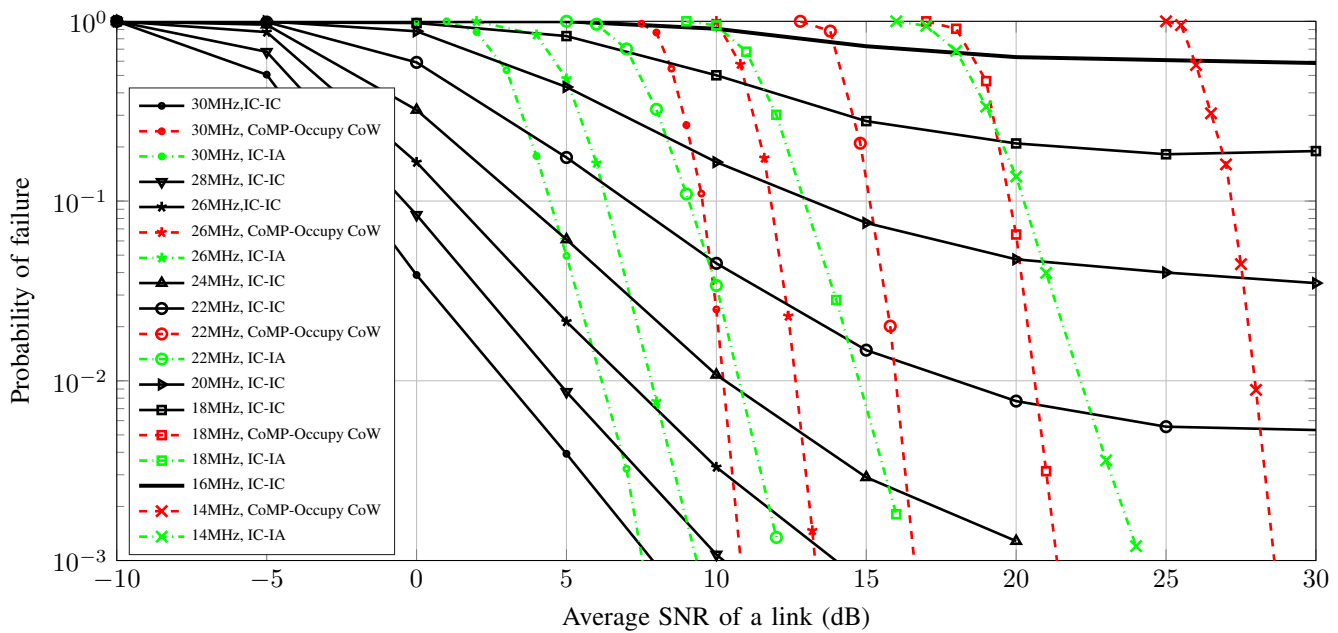
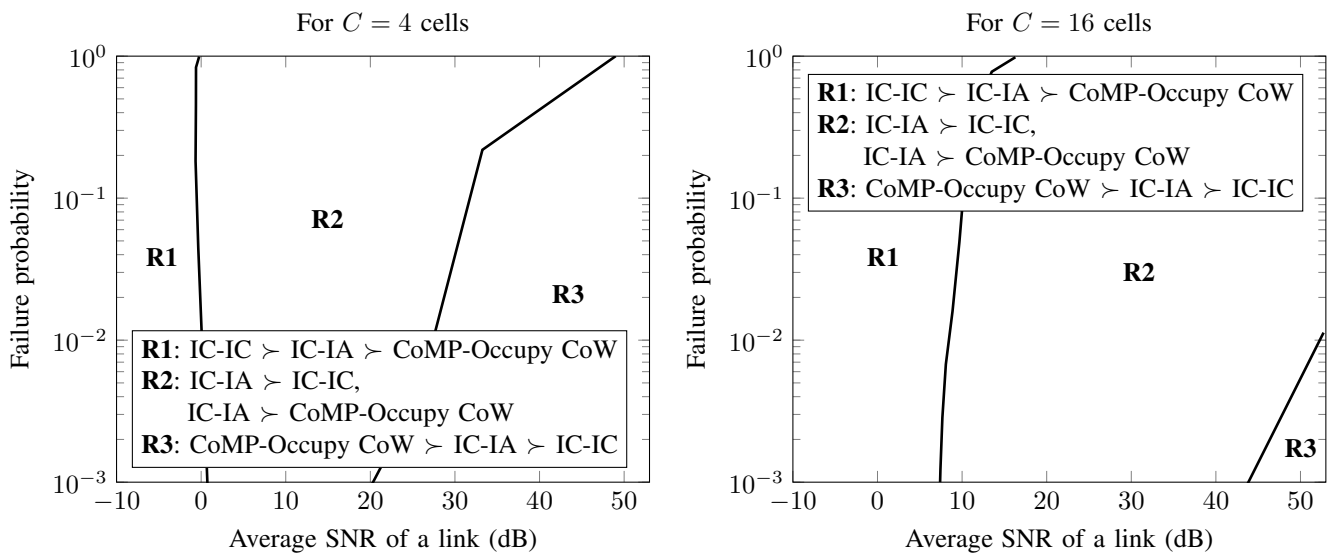
\begin{figure*}[!htbp]
\begin{tabular}{cc}

\begin{tikzpicture}

\begin{axis}[%
scale = 0.8,
width=\columnwidth,
height=0.8\columnwidth,
at={(0in,0in)},
scale only axis,
xmin=-10,
xmax=53,
xlabel={Average SNR of a link (dB)},
ymode=log,
ymin=1e-03,
ymax=1,
yminorticks=true,
ylabel={Failure probability},
axis background/.style={fill=white},
title = {For $C = 4$ cells}
]
\addplot [color=black,solid,line width=1pt]
  table[row sep=crcr]{%
49  1\\
33.25  0.219\\
25.72  0.00411\\
17.538  0.000483\\
};

\addplot [color=black,solid,line width=1pt]
  table[row sep=crcr]{%
-0.237  0.9924\\
-0.607  0.834\\
-0.616  0.453\\
-0.66 0.182\\
-0.34  0.0447\\
0.22  0.00653\\
1  0.0003802\\
};

\end{axis}

\node at (0.6, 3) {\textbf{R1}};
\node at (2.8, 3.5) {\textbf{R2}};
\node at (6, 2.5) {\textbf{R3}};
\node[draw,align=left, fill=white, anchor=south west, execute at begin node=\setlength{\baselineskip}{1.2em}] at (0.2, 0.19) {\textbf{R1}: IC-IC $\succ$ IC-IA $\succ$ CoMP-Occupy CoW\\ \textbf{R2}: IC-IA $\succ$ IC-IC, \\ \ \ \ \ \ IC-IA $\succ$ CoMP-Occupy CoW \\ \textbf{R3}: CoMP-Occupy CoW $\succ$ IC-IA $\succ$ IC-IC};

\end{tikzpicture}

\ \ 

\begin{tikzpicture}

\begin{axis}[%
scale = 0.8,
width=\columnwidth,
height=0.8\columnwidth,
at={(0in,0in)},
scale only axis,
xmin=-10,
xmax=53,
xlabel={Average SNR of a link (dB)},
ymode=log,
ymin=1e-03,
ymax=1,
yminorticks=true,
ylabel={Failure probability},
axis background/.style={fill=white},
title = {For $C = 16$ cells}
]
\addplot [color=black,solid,line width=1pt]
  table[row sep=crcr]{%
42.43  0.00068\\
52.73  0.0113\\
};

\addplot [color=black,solid,line width=1pt]
  table[row sep=crcr]{%
16.3  0.9828\\
13.51  0.777\\
11.61  0.415\\
10.38  0.1552\\
9.68  0.049\\
8.87  0.016\\
8.08  0.00676\\
7.67  0.00285\\
7.44  0.001232\\
7.4  0.001\\
};

\end{axis}


\node at (1.2, 3) {\textbf{R1}};
\node at (4.6, 2.8) {\textbf{R2}};
\node at (6.75, 0.45) {\textbf{R3}};
\node[draw,align=left, fill=white, anchor=south west, execute at begin node=\setlength{\baselineskip}{1.2em}] at (0.2, 3.6) {\textbf{R1}: IC-IC $\succ$ IC-IA $\succ$ CoMP-Occupy CoW\\ \textbf{R2}: IC-IA $\succ$ IC-IC, \\ \ \ \ \ \  IC-IA $\succ$ CoMP-Occupy CoW \\ \textbf{R3}: CoMP-Occupy CoW $\succ$ IC-IA $\succ$ IC-IC};

\end{tikzpicture}
\end{tabular}
\caption{The preference regions for the proposed protocols in failure probability-average SNR plane for $C = 4$ cells (left) and $C = 16$ cells (right), $K = 30$, $b = 160$bits, and $T = 1$ms. Here, ``$\succ$" indicates the preference of protocols for their smaller bandwidth requirement. In low-power region \textbf{R1}, IC-IC requires the smallest bandwidth for achieving a given failure probability. Likewise, in medium-power region \textbf{R2}, the IC-IA, and in the high-power region \textbf{R3} the CoMP-Occupy CoW protocols are preferred. Increasing $C$, expands \textbf{R1} and shrinks \textbf{R3}.}
\label{fig_regions}
\end{figure*}

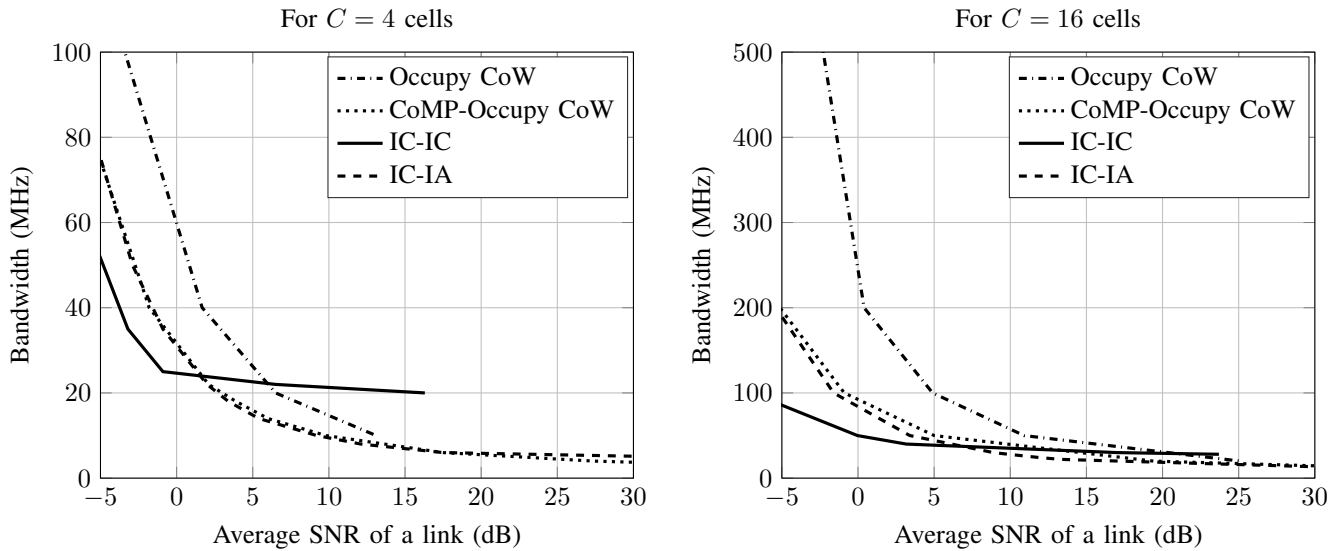
\begin{figure*}[t]
\begin{tabular}{cc}

\begin{tikzpicture}

\begin{axis}[%
scale = 0.8,
width=\columnwidth,
height=0.8\columnwidth,
at={(0in,0in)},
scale only axis,
xmin=-5,
xmax=30,
xlabel={Average SNR of a link (dB)},
xmajorgrids,
ymin=0,
ymax=100,
ylabel={Bandwidth (MHz)},
ymajorgrids,
axis background/.style={fill=white},
title = {For $C = 4$ cells},
legend style={at={(0.99,0.99)},anchor=north east,legend cell align=left,align=left,draw=white!15!black}
]
\addplot [color=black, dashdotted,line width=1.2pt]
  table[row sep=crcr]{%
-9.5000000000000   400.0000000000000\\
  -3.4000000000000   100.0000000000000\\
   1.6800000000000   40.0000000000000\\
 6.5       20\\
   13.1000000000000   10.0000000000000\\
};
\addlegendentry{Occupy CoW};

\addplot [color=black, dotted, line width=1.2pt]
  table[row sep=crcr]{%
-10.1000000000000   200.0000000000000\\
  -7.6100000000000   120.0000000000000\\
  -5.5000000000000   80.0000000000000\\
  -1.8000000000000   40.0000000000000\\
   2.1000000000000   22.0000000000000\\
   3.9000000000000   18.0000000000000\\
   6.0000000000000   14.0000000000000\\
   9.8000000000000   10.0000000000000\\
   17.5000000000000   6.0000000000000\\
   27.4000000000000   4.0000000000000\\
   37.0000000000000   3.0000000000000\\
};
\addlegendentry{CoMP-Occupy CoW};

\addplot [color=black,solid,line width=1.2pt]
  table[row sep=crcr]{%
-10.2000000000000   150.0000000000000\\
  -9.1000000000000   120.0000000000000\\
  -7.0000000000000   75.0000000000000\\
  -5.9000000000000   60.0000000000000\\
  -3.2000000000000   35.0000000000000\\
  -0.9000000000000   25.0000000000000\\
   6.5000000000000   22.0000000000000\\
   16.3000000000000   20.0000000000000\\
};
\addlegendentry{IC-IC};

\addplot [color=black, dashed, line width=1.2pt]
  table[row sep=crcr]{%
-9.4000000000000   200.0000000000000\\
  -7.1100000000000   120.0000000000000\\
  -5.4000000000000   80.0000000000000\\
  -2.9000000000000   50.0000000000000\\
  -0.9000000000000   35.0000000000000\\
   2.0000000000000   22.0000000000000\\
   3.4000000000000   18.0000000000000\\
   5.4000000000000   14.0000000000000\\
   9.2000000000000   10.0000000000000\\
   12.1000000000000   8.0000000000000\\
   17.5000000000000   6.0000000000000\\
   75.7  2\\
};
\addlegendentry{IC-IA};

\end{axis}

\end{tikzpicture}

\ \ 

\begin{tikzpicture}

\begin{axis}[%
scale = 0.8,
width=\columnwidth,
height=0.8\columnwidth,
at={(0in,0in)},
scale only axis,
xmin=-5,
xmax=30,
xlabel={Average SNR of a link (dB)},
xmajorgrids,
ymin=0,
ymax=500,
ylabel={Bandwidth (MHz)},
ymajorgrids,
axis background/.style={fill=white},
title = {For $C = 16$ cells},
legend style={at={(0.99,0.99)},anchor=north east,legend cell align=left,align=left,draw=white!15!black}
]
\addplot [color=black, dashdotted,line width=1.2pt]
  table[row sep=crcr]{%
-8.1  1150\\
0.4  200.0000\\
 4.9  100.0000\\
10.9   50.0000\\
25.1   20.0000\\
};
\addlegendentry{Occupy CoW};

\addplot [color=black, dotted, line width=1.2pt]
  table[row sep=crcr]{%
-7.4400000000000   300.0000000000000\\
  -5.1300000000000   200.0000000000000\\
  -0.9000000000000   100.0000000000000\\
   5.0000000000000   50.0000000000000\\
   19.4000000000000   20.0000000000000\\
   34.7600000000000   12.0000000000000\\
   42.5000000000000   10.0000000000000\\
};
\addlegendentry{CoMP-Occupy CoW};

\addplot [color=black, solid,line width= 1.2pt]
  table[row sep=crcr]{%
-7.0000000000000   100.0000000000000\\
  -0.0400000000000   50.0000000000000\\
   3.1600000000000   40.0000000000000\\
   17.0000000000000   30.0000000000000\\
   23.7000000000000   28.0000000000000\\
};
\addlegendentry{IC-IC};

\addplot [color=black, dashed, line width=1.2pt]
  table[row sep=crcr]{%
-8.9000000000000   400.0000000000000\\
  -5.4000000000000   200.0000000000000\\
  -1.6000000000000   100.0000000000000\\
   3.4000000000000   50.0000000000000\\
   8.8900000000000   30.0000000000000\\
   9.8800000000000   28.0000000000000\\
  12.0000000000000   24.0000000000000\\
  13.3400000000000   22.0000000000000\\
  32.6100000000000   12.0000000000000\\
  42.6400000000000   10.0000000000000\\
};
\addlegendentry{IC-IA};

\end{axis}

\end{tikzpicture}
\end{tabular}
\caption{Tradeoff between the required bandwidth and the required average SNR for having $P_F = 5 \times 10^{-5}$ for $C = 4$ cells (left) and $C = 16$ cells (right). Here, $K = 30$, $b = 160$bits, and $T = 1$ms. The two proposed schemes IC-IC and IC-IA outperform CoMP-Occupy CoW and Occupy CoW at lower transmit powers, corresponding to preference regions \textbf{R1} and \textbf{R2} in Fig~\ref{fig_regions}.}
\label{fig_bwvsp}
\end{figure*}

\subsection{Identifying the Preferable Protocol}
\label{sec_deponw}

In Fig.~\ref{fig_overw16}, for each value of bandwidth, the failure probability curves of IC-IC, IC-IA, and CoMP-Occupy CoW protocols intersect. Following the pattern of intersection points by varying bandwidth, we identify a preference region for each protocol in the failure probability-transmit power plane in Fig.~\ref{fig_regions}. Here, a protocol is preferred over the other, if it achieves a given probability of failure at a smaller bandwidth. We use the preference sign ``$\succ$" to indicate the preference of protocols for their smaller bandwidth requirement.

In Fig.~\ref{fig_regions}, in the low-power region \textbf{R1}, where the required bandwidth for a given failure probability is largest, the IC-IC protocol is preferred over IC-IA, and IC-IA is preferred over the CoMP-Occupy CoW. In the medium-power region \textbf{R2}, the IC-IA protocol outperforms IC-IC and CoMP-Occupy CoW. Finally, in the high-power region \textbf{R3}, where the required bandwidth for a given failure probability is smallest, the CoMP-Occupy CoW protocol is preferred over IC-IA, and IC-IA is preferred over IC-IC.

An important pattern in the preference regions is that by increasing number of cells $C$, the preference regions of the IC-IC and IC-IA protocols expand and the preference region of the CoMP-Occupy CoW protocol shrinks. This illustrates that, by increasing $C$, our proposed protocols improve the bandwidth-power tradeoff for reliability of the CoMP-Occupy CoW protocol without requiring message sharing among controllers, at typical transmit power range.

Fig.~\ref{fig_bwvsp} shows the bandwidth-power tradeoff for achieving $P_F = 5 \times 10^{-5}$ for two values of network size $C = 4$ and $C = 16$. Here, $P_F = 5 \times 10^{-5}$ is the smallest failure probability for which we can run Mont Carlo simulations in a reasonable time. At low transmit powers the IC-IC protocol requires the least amount of bandwidth. This corresponds to region \textbf{R1} in Fig.~\ref{fig_regions}. However, with less bandwidth, the failure probability reaches a floor as a function of SNR. In this regime, the IC-IA protocol avoids the failure probability floor and is the preferred scheme. This corresponds to region \textbf{R2} in Fig.~\ref{fig_regions}. By further reduction in bandwidth and increasing the power, the CoMP-Occupy CoW protocol, which has the highest diversity order, outperforms. This corresponds to region \textbf{R3} in Fig.~\ref{fig_regions}, but we note that the message sharing required by CoMP-Occupy CoW may not always be feasible. As compared to Occupy CoW without message sharing, the proposed protocols, IC-IC and IC-IA, significantly outperform at typical ranges of SNR.


\section{Concluding Remarks}
\label{sec_con}

This paper studies the problem of interference management for ultrareliable low-latency wireless communication in a multiple interfering broadcasts network. The previously proposed Occupy CoW protocol avoids inter-cell interference by orthogonalization, hence, is not scalable in network size. Further, the full reuse of bandwidth in cells while treating all of the inter-cell interference as noise has an adverse effect on the failure probability of the protocol. In this paper, we observe that full bandwidth reuse but with successive decoding and cancellation of as much inter-cell interference as possible can alleviate the dependency of bandwidth-SNR tradeoff on network size. We note that some cell-edge receivers decode inter-cell interference before decoding the intended signal. Based on these observations, we develop two protocols for managing interference and combating fading to meet the stringent requirements of reliability and latency in communication for control applications. When the number of cells in the network increases, the preference regions of our protocols expand and that of the Occupy CoW protocol shrinks.

This paper studies the behavior of the proposed protocols by simulations with a realistic channel model. As a possible direction for future work, it would be useful (but challenging) to analytically characterize the scaling of the required bandwidth for achieving reliability by the proposed interference mitigation protocols with the network size. 
In this work, we study the downlink transmission only. We leave development of interference management schemes for sensors to controllers communication for future studies.







\bibliographystyle{IEEEtran}
\bibliography{IEEEabrv,ref_file}
%




\end{document}